\documentclass[letterpaper,12pt]{article}

\RequirePackage{geometry}
\usepackage{mathptmx}

\usepackage{graphics}

\usepackage[authoryear,comma,sectionbib,sort]{natbib}

\usepackage{placeins}
\usepackage[british]{babel}
\usepackage{dsfont}
\usepackage{bm}
\usepackage[normalem]{ulem}

\newcommand{\bA}{\mbox{\bf A}}
\newcommand{\bB}{\mbox{\bf B}}
\newcommand{\bC}{\mbox{\bf C}}
\newcommand{\bE}{\mbox{\bf E}}
\newcommand{\bG}{\mbox{\bf G}}
\newcommand{\bS}{\mbox{\bf S}}

\newcommand{\bV}{\mbox{\bf V}}
\newcommand{\bv}{\mbox{\bf v}}

\newcommand{\be}{\mbox{\bf e}}
\newcommand{\bK}{\mbox{\bf K}}
\newcommand{\bmu}{\bm{\mu}}
\newcommand{\bSigma}{\bm{\Sigma}}
\newcommand{\graph}{\mbox{\bG=(\bV,\bE)}}
\def\myfrac#1#2{
\hspace{3pt}$\!\!\!^{#1}\!\!$\hspace{1pt}/\hspace{2pt}$\!\!_{#2}\!\!\!$\hspace{3pt}
}
\newcommand{\vI}{\mbox{\textbf{1}}}
\newcommand{\ie}{{\it i.e. }}
\newcommand{\eg}{{\it e.g. }}

\begin{document}
\title{
 Characterization of differentially expressed genes using high-dimensional co-expression networks
 }
\author
{Gabriel Coelho Gon\c{c}alves de Abreu
  \thanks{
          Department of Genetics and Biotechnology,
          Aarhus University.
         }
  \and
 Rodrigo Labouriau$^\ast$
  \thanks{
          Corresponding author: Rodrigo.Labouriau@agrsci.dk
         }
}

\date{}
\maketitle


\maketitle

\begin{abstract}
We present a technique to characterize differentially expressed genes in terms of their position in a high-dimensional co-expression network. The set-up of Gaussian graphical models is used to construct representations of the co-expression network in such a way that redundancy and the propagation of spurious information along the network are avoided. The proposed inference procedure is based on the minimization of the Bayesian Information Criterion (BIC) in the class of decomposable graphical models. This class of models can be used to represent complex relationships and has suitable properties that allow to make effective inference in problems with high degree of complexity (\eg several thousands of genes) and small number of observations (\eg 10-100) as typically occurs in high throughput gene expression studies. Taking advantage of the internal structure of decomposable graphical models, we construct a compact representation of the co-expression network that allows to identify the regions with high concentration of differentially expressed genes. It is argued that differentially expressed genes located in highly interconnected regions of the co-expression network are less informative than differentially expressed genes located in less interconnected regions. Based on that idea, a measure of uncertainty that resembles the notion of relative entropy is proposed. Our methods are illustrated with three publically available data sets on microarray experiments (the larger involving more than 50,000 genes and 64 patients) and a short simulation study.
\end{abstract}

\newpage

\section{Introduction}
The advent of modern functional genomics opened the possibility of studying in details complex relational patterns between genes and phenotypes. The simultaneous access of expression levels of several thousands of genes can be routinely determined by transcriptomic, proteomic and metabolomic technologies \citep{yohannes2008, jourdan2010}. Although a large amount of this type of data has recently been produced, a significant portion of this information might be redundant because gene expression levels tend to be correlated \citep{dorogovtsev2002}. The present work describes techniques to characterize the structure of interdependency between the levels of gene expressions and use this characterization to prune part of this redundancy from large scale gene-expression data (e.g. microarrays and alike techniques). The general idea is to find a simultaneous representation of the values of the expression levels of a large number of genes (co-expression network) and then take advantage of the internal structure of this network to identify groups of genes that are potentially informative.

The following scenario illustrates our ideas. Suppose that the expression levels of several thousands of genes are determined in a number of samples (\eg a microarray experiment). Some of these samples are from diseased and some from healthy tissues. Typically, a list of genes with expression levels presenting statistically significant differences between diseased and healthy tissue is produced. In a naive approach, this list is then directly used to search for patterns of associations between gene regulation and the tissue's disease status. However, since gene expression levels are often correlated
\citep{dorogovtsev2002}, part of those associations might be spurious. For example, there might exist a group of genes, say {\bf \emph{A}}, for which their expression levels are changed in diseased tissues due to the direct effect of the disease. Associations between genes in group {\bf \emph{A}} and the effect of the disease status are genuine. On the other hand, there might exist another group of genes, say {\bf \emph{B}}, for which their expression levels are associated to the expression levels of the genes in group {\bf \emph{A}} due to general gene regulatory mechanisms, but for which the disease has no direct effect. The expression levels of the genes in group {\bf \emph{B}} would be associated with the expression levels of genes in group {\bf \emph{A}} anyway, independently of the disease status of the tissue. In this case, observed associations between the expression levels of the genes in group {\bf \emph{B}} and the disease status of the tissue are spurious. Here the problem is that usually it is not possible to determine if a differentially expressed gene belongs to a group of genes like group {\bf \emph{A}} or group {\bf \emph{B}} and therefore it is not possible in general to establish whether a declared association is spurious or not.

The discussion of the scenario above suggests that it is important to take into account the global structure of the joint expression levels of the genes in an experiment (\ie the structure of the co-expression network) to identify groups of differentially expressed genes that are alike to present a genuine association with the effect of a perturbation (\eg the disease status of a tissue). Associations between a perturbation and  the expression levels of genes cannot be disentangled if the genes are located in a very interconnected region of the co-expression network. Consequently, it is not possible to determine whether observed patterns of differential expression involving those genes are spurious or not. On the other hand, associations involving differentially expressed genes located in less interconnected regions of the co-expression network are less entangled and have a larger potential to represent genuine associations. Another important aspect is that the relative position of a gene in the co-expression network might affect the interpretation of the results. The differential expression of informative genes located in a central area of the network might suggest alterations in basic vital functions. On the other hand, the differential expression of genes located in the periphery of the network might indicate alterations associated to specific aspects related to the perturbation performed. This is a consequence of the flux of information in the co-expression network (see the discussion). In summary, the two aspects of the interpretation of the differential expression of genes discussed above can only be assessed if the position of the differentially expressed genes in the network is taken into account and not only the fact that a gene is differentially expressed or not.

The characterization of a large gene co-expression network is not an easy task since it requires the simultaneous representation of the expression levels of several thousands of genes, possibly forming rather complex patterns of interdependency. We propose here to use the set-up of graphical models \citep{whittaker1990,lauritzen1996}. In this approach the network of co-expression is characterized  by an undirected graph where the vertices (\ie points in a graphical representation) represent the genes in study. Two vertices (\ie two genes)  are connected by an edge (\ie a line in a graphical representation)  when their expression levels are conditionally correlated given the expression levels of all the other genes in the network (see \cite{bollobas2000}, \cite{labouriau2000} and \cite{edwards2010} for similar proposals specific to genetics and gene expression and \cite{whittaker1990}, \cite{lauritzen1996} for the general theory of graphical models). The use of conditional correlations is the first step to avoid that redundant information is represented in the network. However, as it will be clear from the examples presented here, graphical models yield representations of the co-expression network that are still too complex for the purpose of identification of general patterns of differential expression. Therefore we propose an algorithm that uses the internal structure of a rich class of graphical models (the decomposable models) to obtain a tight representation of the co-expression network. This compact representation will yield a better visualization of the distribution of the differentially expressed genes along the co-expression network and will allow us to define a measure of informativeness for differentially expressed genes.

The methods informally described above involve a difficult problem of statistical inference. The type of gene expression data we have in mind typically contains few observations (say 10 to 100) and a much larger number of variables (10 to 60 thousand genes) which renders infeasible the use of standard techniques for statistical inference for graphical models. We will describe a way to circumvent this problem by basing the inference on the minimization of the BIC (Bayesian Information Criterion) and by restricting the class of graphical models to the sub-class of decomposable graphical models. There exists a rather efficient algorithm already implemented for making statistical inference for high dimensional decomposable graphical models (the \emph{gRapHD} package described in \cite{abreu2010} and implemented in R, \cite{R}). This algorithm can also be applied to other similar inference techniques  (\eg the minimization of the AIC (Akaike Information Criterion) or maximization of the entropy) although it can be shown that the minimization of the BIC yields consistent and optimum estimates under rather general conditions (see the discussion).
Moreover, as briefly reported below, we improved the algorithm of minimization of the BIC by implementing a more sophisticated technique to characterize decomposable graphical models which takes advantage of the internal structure of the network. This allows us to treat typical throughput data on gene expression using a reasonable amount of computational resources.

The rest of this work is structured as follows: Section~\ref{sec:methods} briefly revises the basic theory of graphical models, describes an efficient procedure for model inference, places the problem of co-expression networks in the context of graphical models and proposes a general algorithm for clustering genes in a co-expression network. Applications of the proposed  techniques in three publically available data sets and a short simulation study are presented in Section~\ref{sec:applications}. Section~\ref{sec:discussion} presents a brief discussion and some concluding remarks.

\section{Methods}
\label{sec:methods}
\subsection{Co-expression Network viewed as a graphical model}
We construct a gene co-expression network using the framework of graphical models \citep{whittaker1990,lauritzen1996}. In this approach, the expression levels of genes (or probes) are represented by random variables that are the vertices (points) in a graph. More precisely, let $v_1,...,v_p$ be the random variables representing the expression levels of the $p$ genes under study. We will use the terms variable and vertex interchangeably. The network is characterized by the graph $\graph$, where $\bV=\{v_1,...,v_p\}$ is the set of vertices and $\bE$ is a set of unordered pairs of vertices called {\it edges}. Two vertices, $v_i$ and $v_j$, are {\it adjacent} or {\it directly connected} by an edge when the conditional correlation between $v_i$ and $v_j$, given the other vertices, is not zero. In this way, two vertices are adjacent in the graph if, and only if, they carry information on each other that is not already contained in the other vertices. The absence of an edge directly connecting two vertices, say $v_i$ and $v_j$, indicates that the information that $v_i$ carries on $v_j$ (and that $v_j$ carries on $v_i$) is entirely contained in the other vertices of $\bV$. The graph $\graph$ is undirected, \ie if $v_i$ is adjacent to $v_j$ then $v_j$ is also adjacent to $v_i$, since the conditional correlation is symmetric. Here $\graph$ is unweighted, in the sense that there is no meaning attributed to the length of the edges in any graphical representation of $\bG$.

\subsubsection{Basic notions of graph theory}
\label{subsec:graphs}
We introduce below a range of required definitions and notations on graphs. Given an undirected graph $\graph$, a sequence $v_1,...,v_k$ of distinct vertices such that $\{v_i,v_{i+1}\}$ is in $\bE$, for $i=1,...,k-1$ ($k\leq p$), is called a {\it path}. Two vertices, $v_i$ and $v_j$ are {\it connected} when there is a path from $v_i$ to $v_j$. A connected graph has all pairs of vertices connected. We reserve the term {\it directly connected} (or {\it adjacent}) to indicated the existence of an edge between two specific vertices.

A {\it cycle} is a path with equal end points. A connected graph with no cycles is called a {\it tree} and a graph composed of several trees is a {\it forest}. A cycle is {\it chordless} when only successive vertices are adjacent. A graph with no chordless cycles longer than three is said to be {\it triangulated} \citep{bollobas2000}. Clearly, trees and forests are triangulated.

A {\it sub-graph} of $\graph$ is a graph $\bG_A=(\bV_A,\bE_A)$ such that $\bV_A\subseteq\bV$ and $\bE_A\subseteq\bE$. If $\bV = \bV_A$ we call $\bG_A$ a {\it spanning subgraph} of $\bG$. Any spanning subgraph of $\graph$ can be obtained by deleting some of the edges from $\bE$. A spanning subgraph with no cycles is called a {\it spanning tree}. Note that a graph has a spanning tree if, and only if, it is connected. A {\it maximal spanning forest}, or in short a {\it spanning forest}, in $\bG$ is a spanning graph consisting of a spanning tree from each connected component of $\bG$.

A graph $\graph$ is {\it complete} when $\bE$ contains all possible pairs of vertices in $\bV$.  A {\it clique} is a maximally complete sub-graph, in the sense that the addition of any other vertex renders the sub-graph incomplete. A subset $\bC\subseteq\bV$ {\it separates} two disjoint subsets of $\bV$, $\bA$ and $\bB$, if any path from a vertex in $\bA$ to a vertex in $\bB$ contains at least one vertex in $\bC$ \citep{bollobas2000}.

\begin{figure}[!ht]
\centering
	\scalebox{.32}{\includegraphics{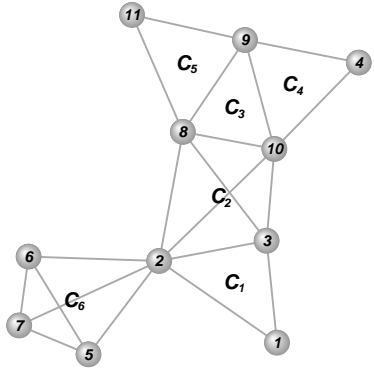}}
\caption{A graph with 11 vertices, 20 edges, and 6 cliques: $C_1=\{1,2,3\}$, $C_2=\{2,3,8,10\}$, $C_3=\{8,9,10\}$, $C_4=\{4,9,10\}$, $C_5=\{8,9,11\}$, and $C_6=\{2,5,6,7\}$.}
\label{fig:example}
\end{figure}

We introduce next a convenient representation of any triangulated graph which will be crucial for describing the statistical inference and the reduction algorithm proposed. Let $\graph$ be a triangulated graph and $C_1,\dots,C_k$ be an enumeration of all the cliques of $\bG$. Consider the sets $S_1, \dots , S_k$ given by,  $S_1 = \emptyset$ and for $ j=2,\dots,k$,
$$
 S_j = C_j\cap\left(\bigcup_{i=1}^{j-1} C_i\right) \, .
$$
The sets $S_1,\dots,S_k$ are called the {\it separators} of the sequence of cliques. That is, a separator $S_j$ is given by the intersection of $C_j$ with all previous cliques $C_i$, $i<j$.


According to the classic theory of graphs, the graph $\bG$ is triangulated if, and only if, there exist an enumeration of its cliques such that each of the separators $S_1, \dots ,S_k$ are complete and, for each $i$ in $\{ 1, \dots , k\}$ there is a $j<i$ such that $S_i \subseteq C_j$ \citep{golumbic1980,lauritzen1996}.
One such enumeration is called a {\it perfect sequence of cliques} and will be used to perform efficient recursive calculations necessary in inference procedures involving triangulated graphs.
There are several efficient algorithms to find in linear time a perfect sequence of a triangulated graph, as the maximum cardinality search \citep{tarjan1984} and the lexographic search \citep{rose1976}. For example, in the graph represented in Figure~\ref{fig:example}~(A) one perfect sequence is: $C_1=\{1,2,3\}$, $C_2=\{2,3,8,10\}$, $C_3=\{8,9,10\}$, $C_4=\{4,9,10\}$, $C_5=\{8,9,11\}$, and $C_6=\{2,5,6,7\}$, with the respective separators $S_1=\emptyset$, $S_2=\{2,3\}$, $S_3=\{8,10\}$, $S_4=\{9,10\}$, $S_5=\{8,9\}$, $S_6=\{2\}$.

\subsubsection{Statistical inference of networks}
\label{subsec:basic}
It is assumed that the observations are independent realizations of a multivariate normal distribution. Let $\bv_1, \dots , \bv_n$ be $n$ independent observations representing the expression levels of $p$ genes and assume that
$$
 \bv_i \sim N_p \left ( \bmu , \bSigma \right), \mbox{ for } i=1,\dots ,
 n \, ,
$$
where $N_p (\, \cdot \, , \, \cdot \,)$ represents the $p$-dimensional multivariate normal distribution. Here the covariance matrix $\bSigma$ is understood as the parameter of interest and the mean vector $\bmu$ is considered as a nuisance parameter, since our main concern is to characterize the covariance structure. If we consider a subset of the $p$ variables in $\bv_i$, $\bv_i^k=\{v_i^1,...,v_i^k\}$, with $k\leq p$ (\ie if $k=p$ then $\bv_i^k=\bv_i$), then the marginal density function of $\bv_i^k$ is given by
$$
f(\bv_i^k)=\frac{1}{(2\pi)^{k/2}[\det{(\bSigma_k)}]^{1/2}}e^{-\frac{1}{2}(\bv_i^k-\bmu_k)'\bSigma_k^{-1}(\bv_i^k-\bmu_k)},
$$
where $\bSigma_k$ and $\bmu_k$ are respectively the covariance matrix and mean vector of $\bv_i^k$.

A graphical model is called {\it decomposable} if the related graph is triangulated. The likelihood function of any decomposable graphical model can be factorized according to its cliques and separators. This greatly simplifies the calculation of the quantities related to the likelihood function. For instance, the log-likelihood function can then be written as
$$
l(\bSigma,\bmu|\bv)=\sum^n_{i=1}\sum_{C_j\in\cal{C}}\ln f\left(\bv_i^{C_j}\right) -\sum^n_{i=1}\sum_{S_j\in\cal{S}}\nu(S_j)\ln f\left(\bv_i^{S_j}\right),
$$
where $\bv=\{\bv_1,...,\bv_n\}$, $\cal{C}$ and $\cal{S}$ are respectively the sets of cliques and separators of $\bG$, $C_j$ is a perfect enumeration of the cliques in $\cal{C}$ with respective separators $S_j$, and $\nu(S_j)$ is the multiplicity of the separator $S_j$ in $\cal{S}$, $j=1,...,k$ \citep{lauritzen1996}.
The Bayesian Information Criterion, $\mbox{BIC}(\bG)$, can then be calculated as
$$
\mbox{BIC}(\bG)=-2\, l(\bSigma,\bmu|\bv)+\kappa\ln(n),
$$
where $\kappa$ is the number of estimated parameters in the model, given by $2p+|\bE|$ and $|\bE|$ is the cardinality of the set of edges.

The classic method for likelihood-based statistical inference under Gaussian graphical models (covariance selection models) involves the inversion of the sample covariance matrix. This technique cannot be directly applied here because typically gene expression data contains a much larger number of variables (the genes, $p$) than observations (individuals or samples, $n$), implying that the empirical covariance matrix is singular \citep{dykstra1970}. For this reason we propose to infer the co-expression network by finding the decomposable graphical model with minimum BIC. This technique yields consistent and optimum estimates \citep[see][]{haughton1988}, although our methods could easily be adapted to other similar inference techniques  (\eg the minimization of the AIC (Akaike Information Criterion) or maximization of the entropy).

The search for a graphical model with minimum BIC is made in two steps: First the spanning forest with minimum BIC is found and then a forward search is performed by successively adding edges that reduce the BIC and still preserve the decomposability of the graph.
The spanning forest with minimum BIC is constructed in the following way. Starting with a graphical model containing no edges, the edges that decrease mostly the BIC among those edges whose addition does not generate cycles in the graph are successively added. Here the improvement in the BIC due to the addition of the edge $\{v_i,v_j\}$, $I_{\{v_i,v_j\}}$, is given by:
$$
I_{\{v_i,v_j\}}=-2\ln\left\{\frac{\prod_{i=1}^nf(\bv^{A}_i)f(\bv^{B}_i)/f(\bv^{S}_i)}{\prod_{i=1}^nf(\bv^{A\cup S\cup B}_i)}\right\}+\kappa\ln(n),
\label{eq:fact_BIC}
$$
where $v_i\in\bA$, $v_j\in\bB$, and $\bS$ separates $\bA$ and $\bB$ in the current graph. The sets $\bA$ and $\bB$ are cliques in the current graph. The constant $\kappa$ is the number of free parameters estimated, and $n$ is the number of observations (number of arrays) \citep{edwards2010}.
The algorithm is:
\begin{list}{\labelitemi}{}
\setlength{\itemsep}{-1pt}
	\item[\textit{Forest search}]
	\item[{\bf begin}]
	\item[1.] $\bV = \{v_1,\dots,v_p\}$ and $\bE = \emptyset$
	\item[2.] $\graph$
	\item[3.] {\bf for every} $\{v_i,v_j\}$, $i\neq j$
	\item[4.] \hspace{5pt} calculate $I_{\{v_i,v_j\}}$ and store in $\Gamma$
	\item[5.] {\bf while} $|\Gamma|>0$ and $\bG$ is a forest
	\item[6.] \hspace{5pt} select $\be=\{v_i,v_j\}$ with minimum $I_{\{v_i,v_j\}}$ in $\Gamma$
	\item[7.] \hspace{5pt} {\bf if} $I_{\{v_i,v_j\}}<0$ and $\bE\cup\be$ does not generate a cycle
	\item[8.] \hspace{15pt} add $\be$ to $\bE$
	\item[9.] \hspace{5pt} remove $\be$ from $\Gamma$
	\item[{\bf end}]
\end{list}

Once the spanning forest with minimum BIC is found the search continues by successively adding edges that further decrease the BIC, as described in the algorithm below. The algorithm takes advantage of the fact that the addition of an edge changes the graph locally, thus only the values of $I_{\{ \, \cdot \, , \, \cdot \, \}}$  for vertices in the altered region need to be re-calculated. Denote by $\cal{E} (\bG)$ the class of edges that can be added to a decomposable graph $\bG$ preserving its decomposability. The algorithm is the following:
\newpage
\begin{list}{\labelitemi}{}
\setlength{\itemsep}{-1pt}
	\item[\textit{Decomposable search}]
	\item[{\bf begin}]
	\item[1.] $\bV = \{v_1,\dots,v_p\}$
	\item[2.] $\bG=\{\bV,\bE\}$
	\item[3.] stop = false
	\item[4.] {\bf while} not stop
	\item[5.] \hspace{5pt} find $\cal{E}(\bG)$
	\item[6.] \hspace{5pt} update the values of $I_{\{v_i,v_j\}}$
	\item[7.] \hspace{5pt} select $\be=\{v_i,v_j\}$ with minimum $I_{\{v_i,v_j\}}$
	\item[8.] \hspace{5pt} {\bf if} $I_{\{v_i,v_j\}}>0$
	\item[9.] \hspace{15pt} stop = true
	\item[10.] \hspace{5pt} {\bf else}
	\item[11.] \hspace{15pt} add $\be$ to $\bE$
	\item[{\bf end}]
\end{list}

\noindent
The step 5 in the algorithm above - find $\cal{E}(\bG)$ - is typically rather time consuming if implemented as in \cite{abreu2010}. Therefore we used the technique suggested by \cite{deshpande2001}:  Define the {\it clique graph} of a decomposable graph $\graph$ as a new graph $\bG^{cg}$, in which the vertices are the cliques of $\bG$, and two vertices, $v^{cg}_i$ and $v^{cg}_j$ (representing cliques $C_i$ and $C_j$ of $\bG$), are directly connected by an edge if and only if the set $C_i\cap C_j$ separates $C_i\backslash(C_i\cap C_j)$ and $C_j\backslash(C_i\cap C_j)$ in $\bG$. An edge $\{v_a,v_b\}$ is in $\cal{E}(\bG)$ if, and only if, $v_a\in C_i$ and $v_b\in C_j$, given that the edge $\{C_i,C_j\}$ is in $\bG^{cg}$.

\subsection{An algorithm for clustering genes}\label{section:clustering}

We describe below a method to construct a compact representation of the distribution of differentially expressed genes along a complex co-expression network. We presuppose that the genes in study can be classified in two categories: differentially expressed (DEG) and non-differentially expressed genes (NDEG).

Suppose that the graph $\graph$, representing the co-expression network, can be split in $k$ connected sub-graphs, say $C_1, \dots , C_k$, termed the components of $\bG$. Each of these components can be classified as: differentially expressed gene dense (DEGD) or non-differentially expressed gene dense (NDEGD); if the proportion of DEG in the component exceeds or not a pre-fixed threshold $\alpha$, respectively. Here the proportion of DEG in $\bV$ is a natural choice for $\alpha$. The idea of the algorithm below is to merge the components that are neighbours in the graph $\bG$ and have the same classification, producing in this way a collection of subsets of $\bV$ which elements are called {\it gene clusters}. We can then construct a second order graph $\bG_K = \left (\bK, \bE_K \right)$ for which the vertices are the gene clusters and two clusters, $K_1$ and $K_2$, are adjacent when there exist two genes $v_1\in K_1$ and $v_2\in K_2$ such that $v_1$ and $v_2$ are adjacent in $\graph$. The graph $\bG_K = \left (\bK, \bE_K \right)$ is called the {\it cluster graph}.
When the representation of the co-expression network is a triangulated graph,  it is natural to construct the graph components using the cliques.

Let $\graph$ be a triangulated graph representing the co-expression network for which the vertices can be classified as DEG or NDEG. The algorithm below will find the gene clusters and output the associated cluster graph:
\begin{list}{\labelitemi}{}
\setlength{\itemsep}{-1pt}
	\item[\textit{Clustering}]
	\item[{\bf begin}]
	\item[1.] $\alpha\leftarrow$ proportion of DEG in $\bV$
	\item[2.] find all cliques $C_1,\dots,C_k$ of $\graph$
	\item[3.] {\bf for} each $C_i$, $i=1,\dots,k$
	\item[4.] \hspace{7pt} $\alpha_i\leftarrow$ proportion of DEG in $C_i$
	\item[5.] \hspace{7pt} classify $C_i$ as DEGD if $\alpha_i\geq\alpha$
	\item[6.] \hspace{7pt} otherwise classify $C_i$ as NDEGD
	\item[7.] {\bf for} $j$ in $\left\{ \mbox{DEGD}, \mbox{NDEGD} \right\}$
	\item[8.] \hspace{7pt} $\bG_j\leftarrow$ sub-graph of $\graph$ for all vertices in $j$-classified cliques
	\item[9.] \hspace{7pt} each connected component of $\bG_j$ is a $j$-dense cluster
  \item[10.] Form the cluster graph  $\bG_K = \left (\bK, \bE_K \right)$
	\item[{\bf end}]
\end{list}

\begin{figure}[!ht]
\centering
	\scalebox{.22}{\includegraphics{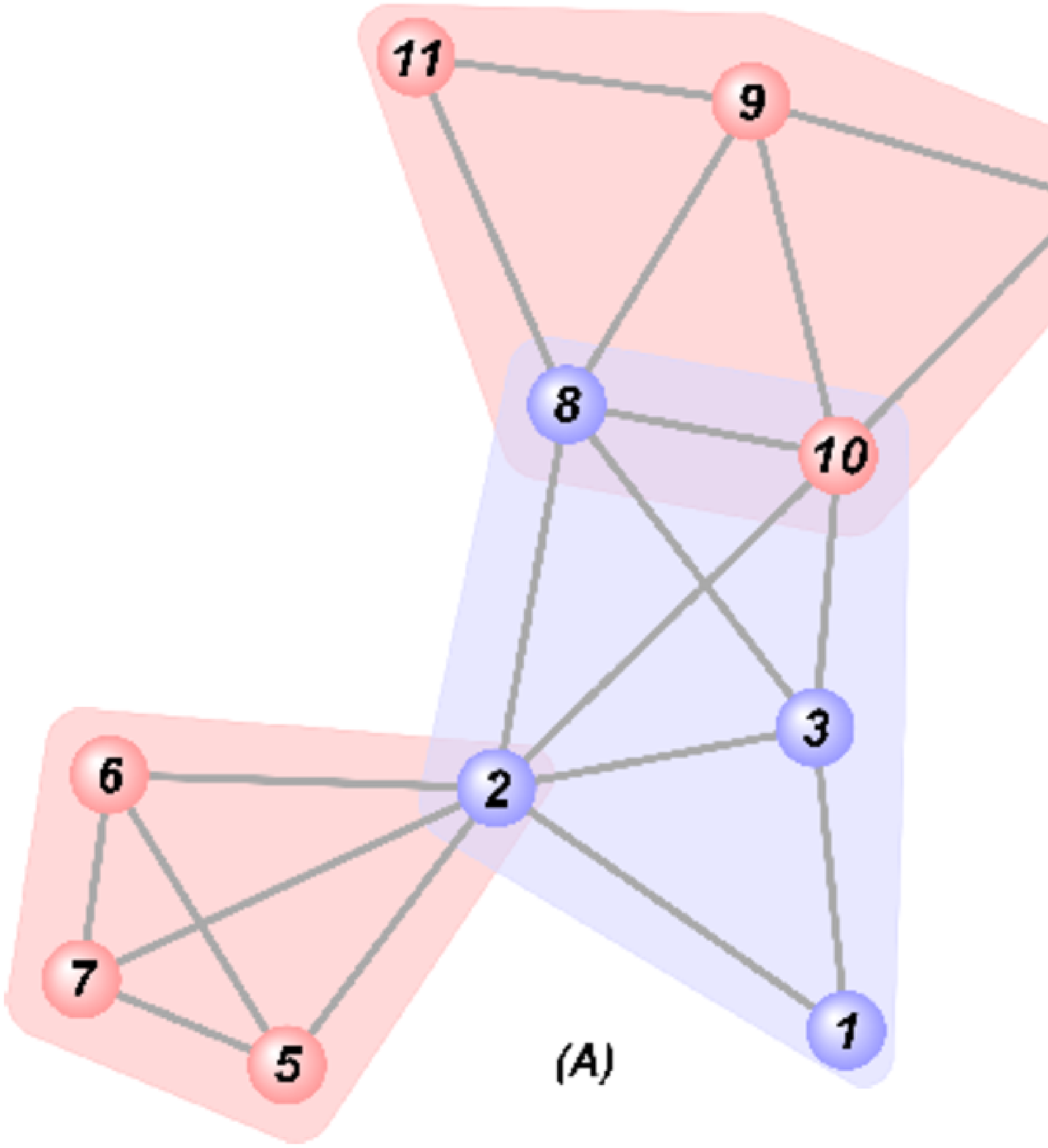}} \hspace{2cm}
	\scalebox{.32}{\includegraphics{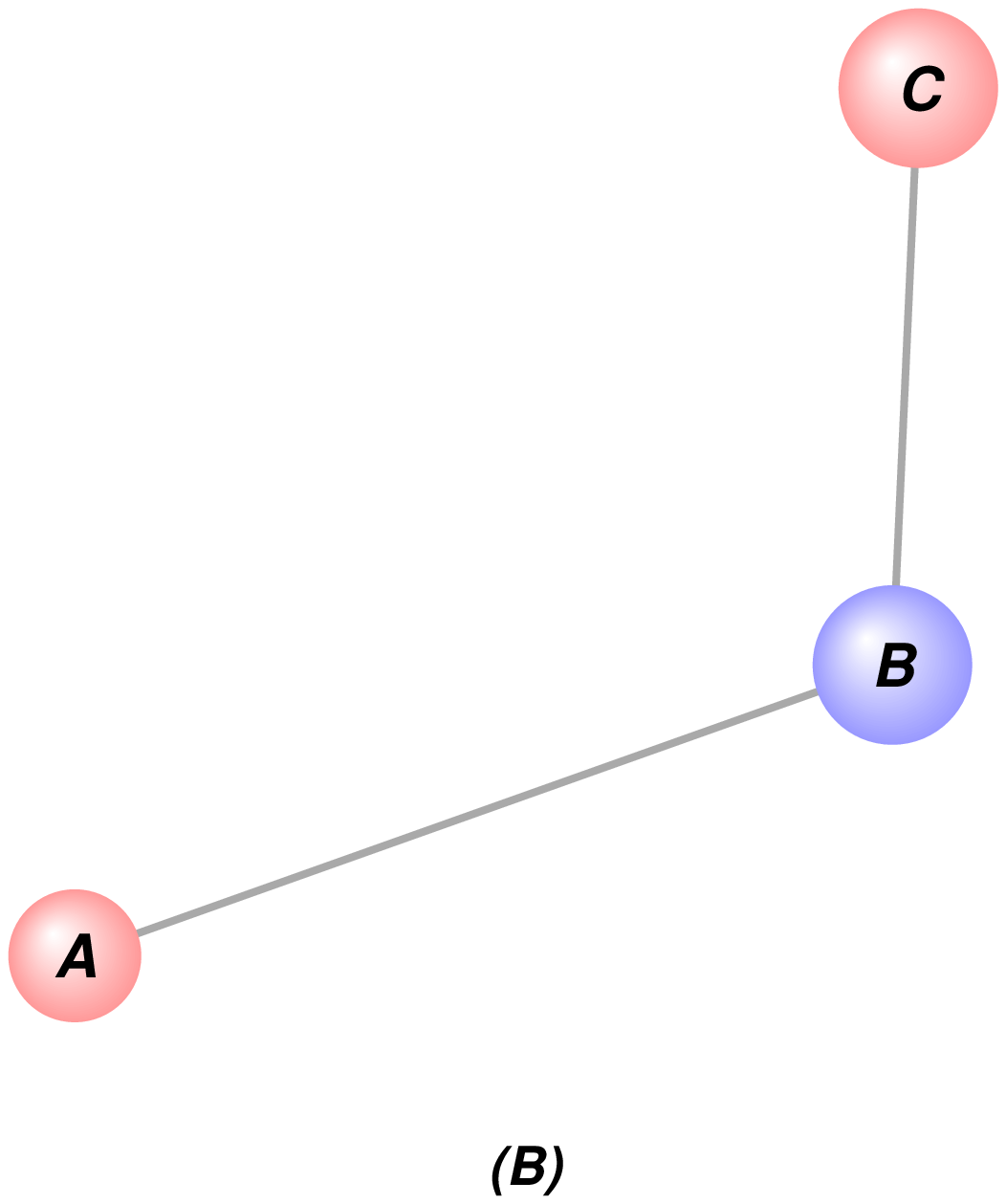}}
\caption{(A) A graph representing DEG (in red) and NDEG (in blue); (B) The cluster graph of (A), where each dot now represents a gene cluster (DEGD in red and NDEGD in blue).}
\label{fig:clustProc}
\end{figure}

Figure~\ref{fig:clustProc}~(A) shows an example with the two distinct groups of vertices: DEG containing the genes labelled $5,6,7,9,10,11$ (represented in red) and NDEG containing the genes labelled $1,2,3,4,8\}$, (represented in blue). The graph is triangulated with cliques given by $C_1=\{1,2,3\}$, $C_2=\{2,3,8,10\}$, $C_3=\{8,9,10\}$, $C_4=\{4,9,10\}$, $C_5=\{8,9,11\}$, and $C_6=\{2,5,6,7\}$. If the threshold $\alpha$ is \myfrac{6}{11}, the cliques $C_3$, $C_4$, $C_5$, and $C_6$ are classified as DEGD. As $C_6$ is not directly connected to other DEGD clique, it solely composes the cluster $A=\{2,5,6,7\}$. The cliques $C_1$ and $C_2$ are directly connected, forming the cluster $B=\{1,2,3,8,10\}$. The cluster $C=\{4,8,9,10,11\} $ is formed by the cliques $C_3$, $C_4$, and $C_5$, all directly connected and DEGD. Solving the separators, the final clusters are $A=\{2,5,6,7\}$, $B=\{1,3,10\}$, and $C=\{4,8,9,11\}$. Having the vertices clustered in this way, the graph can be drastically reduced to the  cluster graph displayed in Figure~\ref{fig:clustProc}~(B).

\subsubsection{A measure of the level of uncertainty of a gene}
The following basic idea will be used to construct a measure of the level of uncertainty of a gene. Suppose that we establish that a gene, $v$ is differentially expressed. We argue that if $v$ is located in a highly interconnected region of the network containing many other differentially expressed genes, then to say that $v$ is differentially expressed essentially does not reduce the uncertainty that we had before the observation is done. This occurs because the expression levels of many other differentially expressed genes are associated to the expression level of $v$. On the other hand, if the interconnected region of the network where $v$ is located contains few (or in the extreme case, none) other differentially expressed genes, then declaring $v$ differentially expressed reduces significantly the uncertainty that we had before observing the data; thus suggesting that the informational content of such declaration is high.

The starting point of the construction below is a segmentation of the network. We assume that the genes of the network are classified in disjoint subsets of genes constructed in such a way that the expression levels of the genes located in the same subset are associated. One way to construct such a stratification is by joining connected cliques in a decomposable graphical model as proposed in section \ref{section:clustering}.
Since the gene clusters produced there are not necessarily disjoint, we adopt the following convention: if a vertex $v$ is in the intersection of two gene clusters, then $v$ is moved to the cluster that has the opposite classification (with an arbitrary choice in the ambiguous cases). This choice minimizes the number of clusters falsely classified as DEGD. A reciprocal convention would maximize the number of cliques truly classified as DEGD.

The algorithm defined there joined neighbouring cliques with predominance of differentially expressed genes and neighbouring cliques with predominance of non-differentially expressed genes. %
The proposed measure of uncertainty for a cluster ${K}$ is given by
$$
 \rho_0 = \rho_0 ({K}) =
 - \frac{\eta_{K} }{p}
 \log \left ( \frac{\eta_{K} }{p}  \right) \,\, ,
$$
where $\eta_{K}$ is the number of differentially expressed genes in ${K}$ and $p$ is the total number of genes in the network. The measure $\rho_0$ resembles the classic entropy measure of uncertainty, the larger is $\rho_0$ the larger is the uncertainty. Since $\rho_0({K})$ is a relative measure of uncertainty, it is standardized by dividing $\rho_0({K})$ by the larger  $\rho_0$ found in the network in study, say $\rho_{0;max}$, defining in this way the a relative measure of uncertainty by
$$\rho = \rho ({K}) =
 - \frac{ \rho_0 ({K})}{\rho_{0;max}} \,\, .
$$
We attribute the measure of uncertainty $\rho_0 ({K})$ to each differentially expressed gene in the cluster ${K}$.

\section{Applications}
\label{sec:applications}

\subsection{Three examples}
In order to illustrate the proposed methods we present three examples representing rather different situations.
In the first example - human healthy and diseased gingival tissues - we study a relatively large network (54,675 vertices). This example illustrates how difficult it might be to identify patterns in the distribution of differentially expressed genes in large co-expression networks by directly  examining a decomposable graphical model. However, applying the proposed clustering procedure allowed us to identify regions of higher concentration of differentially expressed genes and to visualize general patterns.

The second example - muscle water holding capacity in pigs - illustrates a situation were most of the differentially expressed genes are located in less interconnected peripheral regions of the network and just a small proportion of the differentially expressed genes presented high uncertainty. An opposite situation occurs in the third example - intra muscular fatness in pigs - where most of the differentially expressed genes are located in a large central cluster and just few genes had low uncertainty. The first example occupies an intermediary position, presenting differentially expressed genes with high, intermediate and low uncertainty.

\ \linebreak
\noindent {\bf Example 1 - Human healthy and diseased gingival tissues}\\

\noindent
This example arises from a microarray experiment analysed and described in details in \citet{demmer2008}.  In this study, biopsies from diseased and healthy gingival tissues were collected from 90 patients and hybridized to individual arrays (Human Genome U133 Plus 2.0, Affymetrix) with 54,675 probe sets in each array. The data is publicly available at the Gene Expression Omnibus (accession GSE10334).

Here we selected the sub-sample of the 64 patients for which one biopsy of healthy tissue and at least two biopsies of diseased tissues were available. This yielded a data set with 194 arrays (130 from diseased sites). Next, each probe was classified as differentially expressed or not using the following procedure: The expression of each probe was studied using a gaussian linear mixed model containing a random component taking the same value for each patient and a fixed effect representing the tissue status (healthy or diseased). The significance of the fixed effect was tested by comparing the likelihood ratio statistics to the quantiles of the empiric distribution of 1,000 parametric bootstrap samples drawn under the null hypothesis (absence of effect of tissue status). After applying a false discovery ratio  correction for multiple comparisons the probes were classified as differentially expressed (DEP), \ie presenting statistically significant effect of the tissue status ($p<0.05$), or not differentially expressed (NDEP).  We found 23,773 differentially expressed probes (\ie around 43\%).

\begin{figure}[!ht]
 \centering
 \scalebox{.15}{\includegraphics{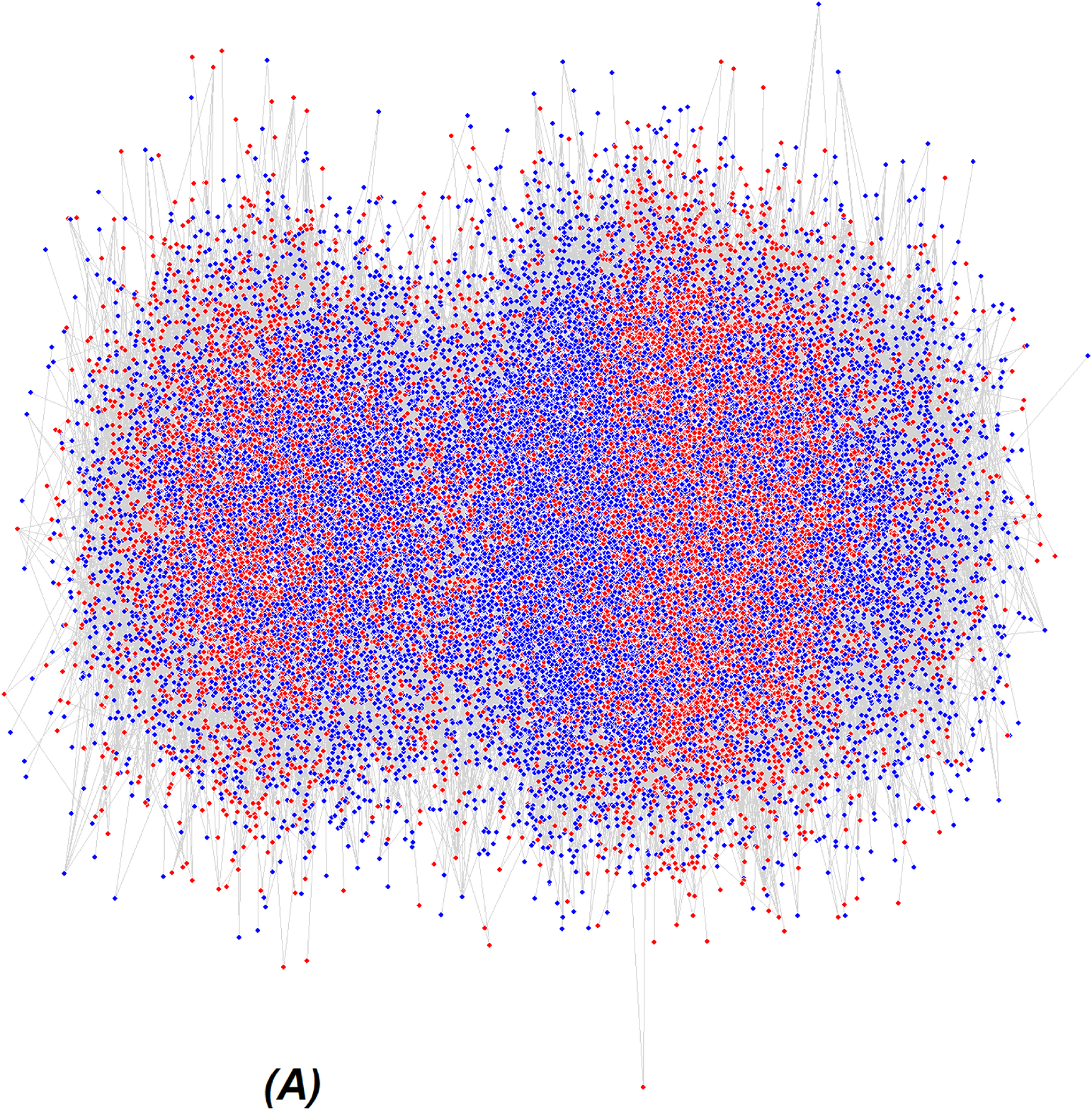}}
 \scalebox{.55}{\includegraphics{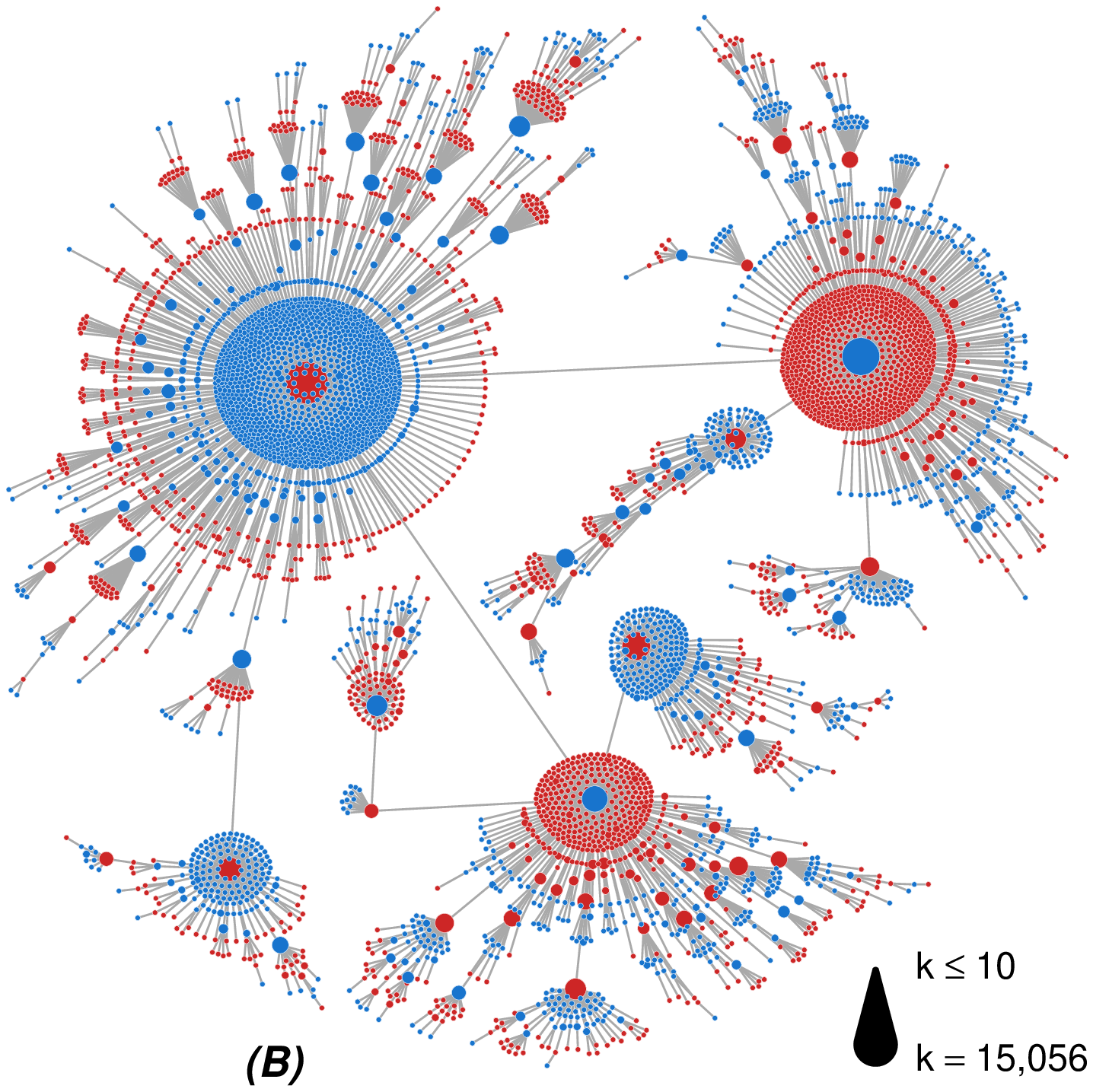}}
 \caption{Co-expression network of Example 1. (A) Raw representation of the decomposable graphical model with minimum BIC; red and blue points represent differentially expressed and non-differentially expressed probes, respectively.
 (B) Network representation obtained by the clustering procedure; each point represents a cluster, which size is proportional to the number of probes in the cluster;
 clusters with predominance of differentially expressed probes are marked in red and the others in blue.
}
\label{fig:perio54675a}
\end{figure}

\begin{figure}[!ht]
 \centering
 \scalebox{.45}{\includegraphics{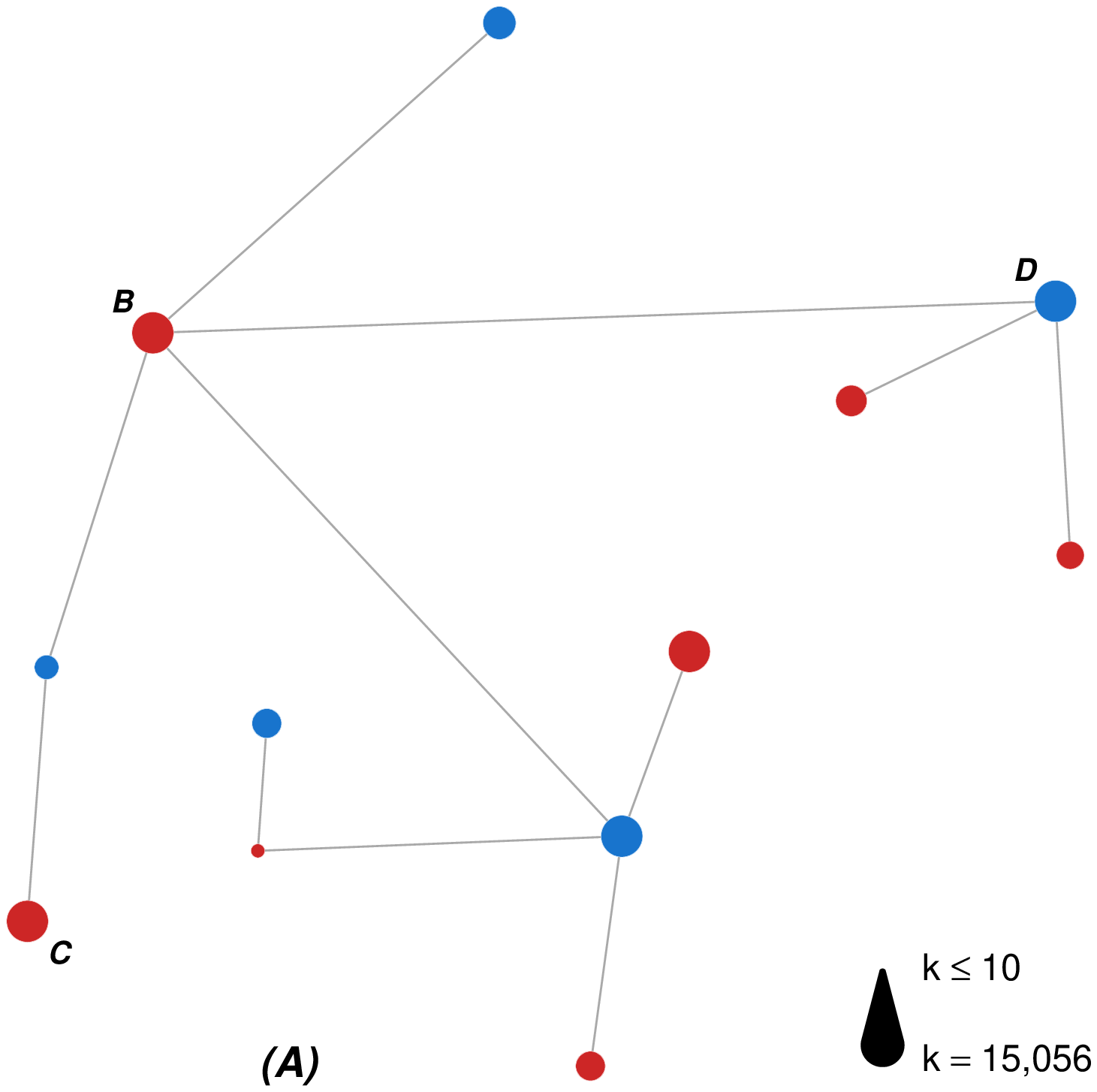}}
 \scalebox{.45}{\includegraphics{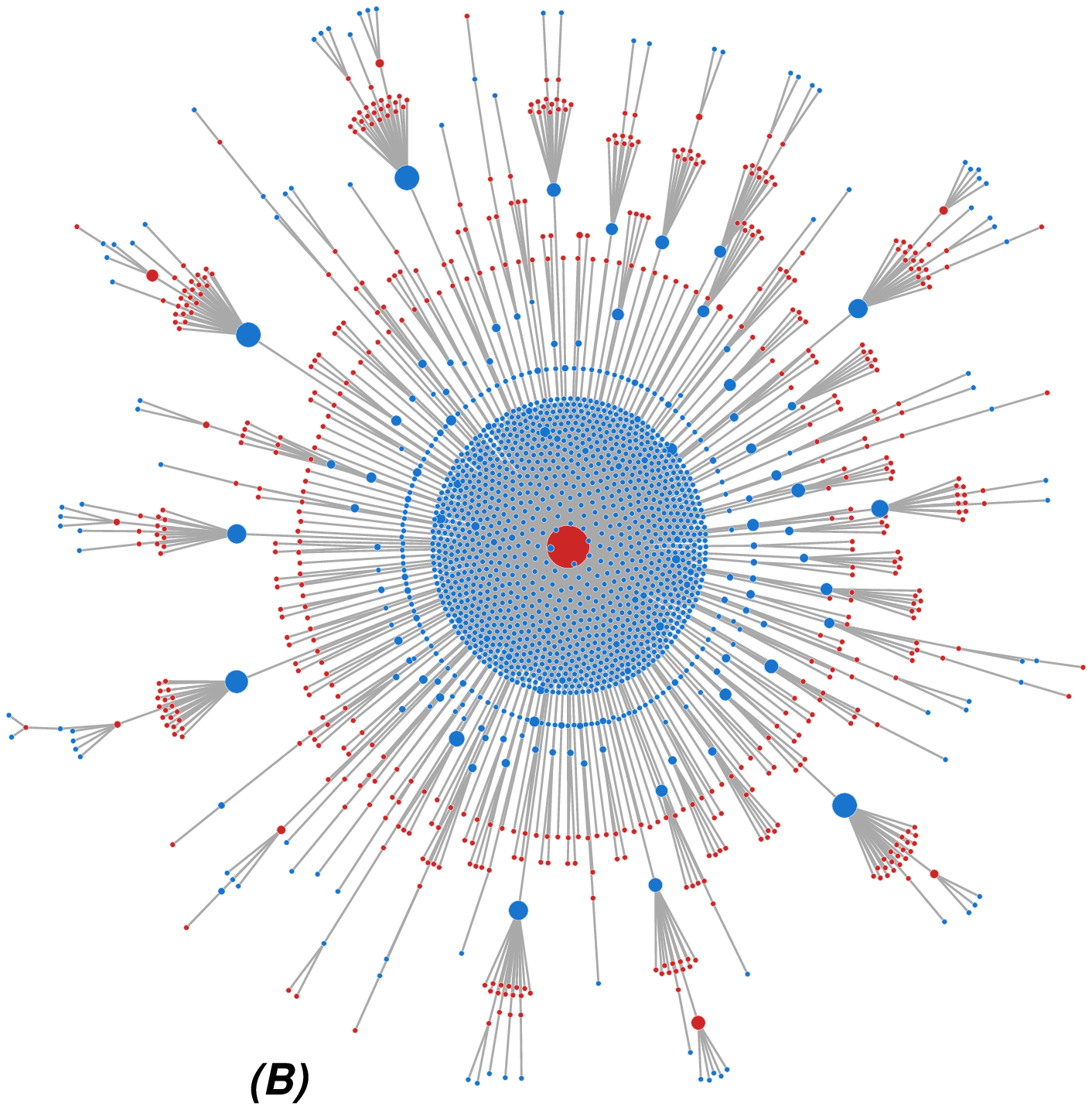}}
 \scalebox{.45}{\includegraphics{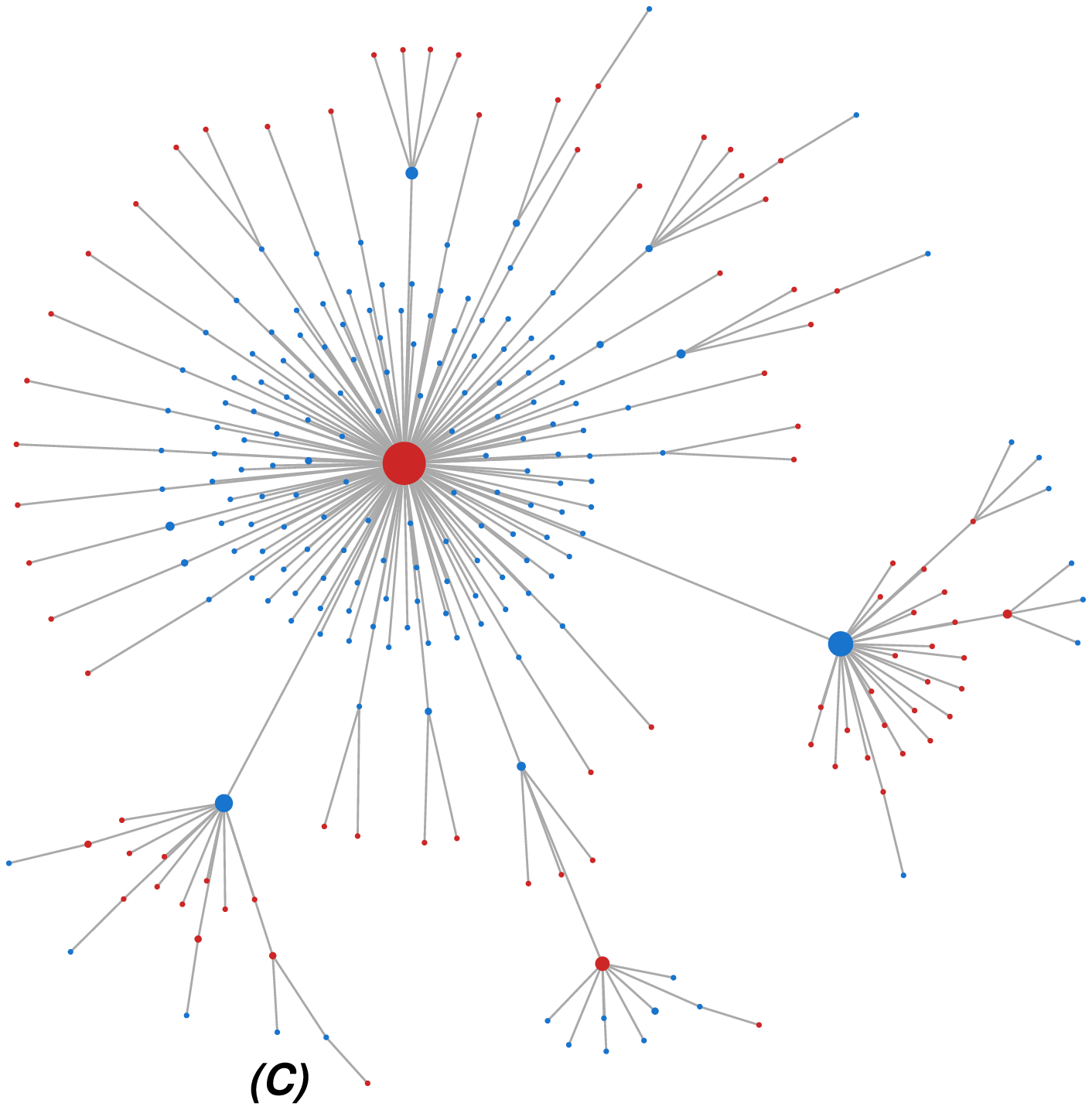}}
 \scalebox{.45}{\includegraphics{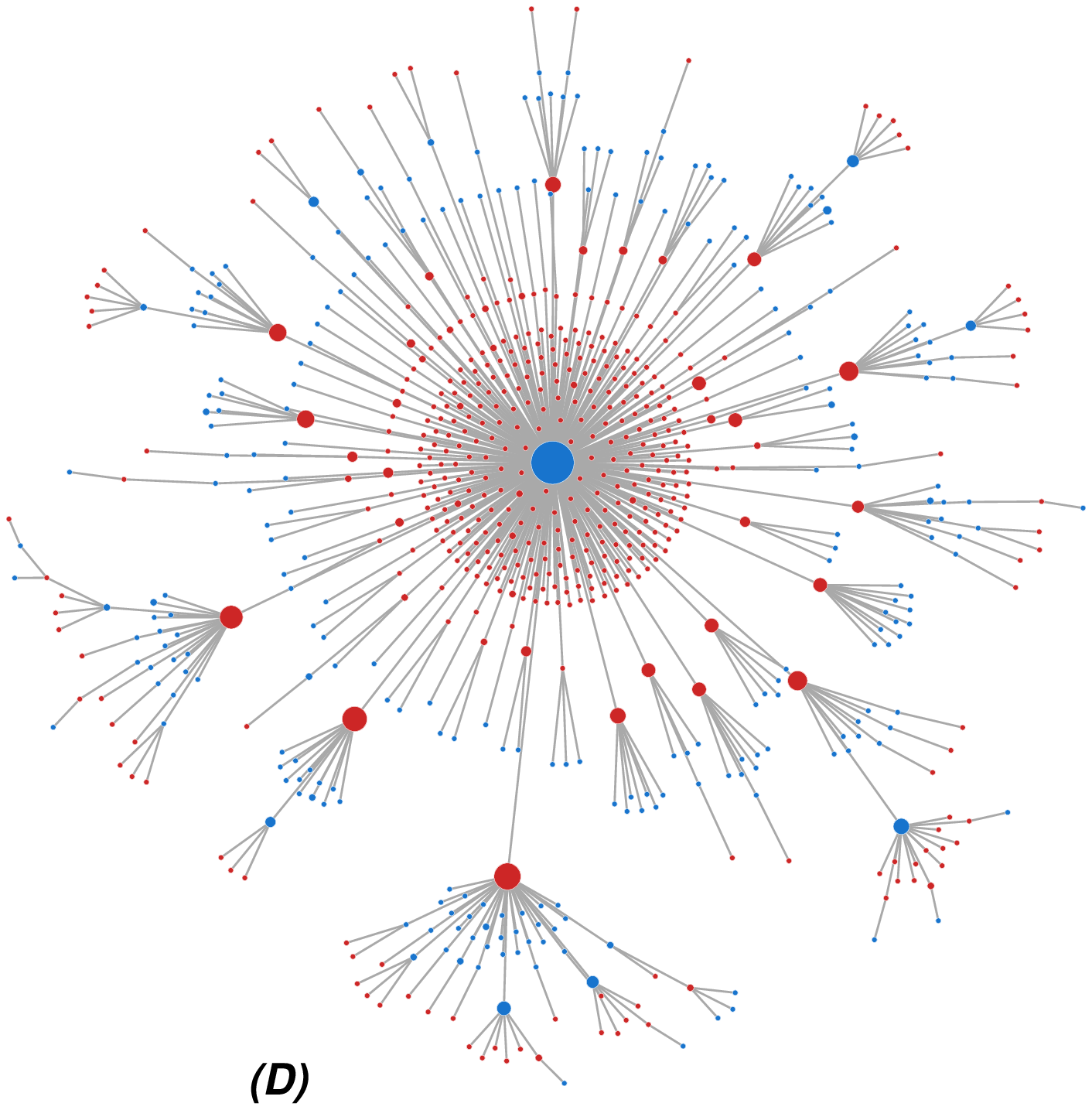}}
 \caption{Details on the co-expression network of Example 1. (A) The smallest tree connecting the ten largest clusters; the sizes of the points are proportional to the number of probes in the corresponding clusters; clusters with predominance of differentially expressed probes are marked in red and the others in blue. (B) - (D) Detailed view of the three largest clusters designated by the respective letters in (A).}
\label{fig:perio54675b}
\end{figure}

The co-expression network was inferred using the data of healthy sites. The spanning forest with minimum BIC presented only one connected component. The forward search for a decomposable graphical model resulted in the addition of 24,735 edges, producing a network where 37\% of the vertices were leaves (\ie vertices with degree 1) and the maximum vertex's degree was 131.
Figure~\ref{fig:perio54675a}~(A) displays the inferred network. It is not possible to identify any pattern in this representation of the network because the complexity of the inferred decomposable graphical model is too high. On the other hand, this complexity was drastically reduced by applying the clustering procedure, as showed in Figure~\ref{fig:perio54675a}~(B). We found 5,081 clusters among which 2,378 (\ie around 47\%) were classified as containing predominantly DEP. Figure~\ref{fig:perio54675b}~(A) displays the smallest tree containing the ten largest clusters and Figures~\ref{fig:perio54675b}~(B)-(D) show three of those clusters with their closest neighbouring clusters.

\newpage

\FloatBarrier

The largest cluster encountered (see Figure~\ref{fig:perio54675b}~(B)) contained 15,056 probes (around 28\% of the total number of probes) and occupied a central position in the network (see Figure~\ref{fig:perio54675b}~(A)). The level of uncertainty of the DEP in this cluster is high, since the cluster contained 44\% of the DEP. However, examining Figure~\ref{fig:perio54675a}~(B) and the distribution of the uncertainty index $\rho$ (see Figure~\ref{fig:rho}), it turns out that the network contained also DEP with intermediate and low levels of uncertainty.

\ \linebreak
\noindent {\bf Example 2 - Muscle water holding capacity in pigs}\\

\noindent
This example stems from a study on muscle water holding capacity (WHC) in pigs reported by \citet{ponsuksili2008}. The global pattern of gene transcription, assessed by Affimetrix Porcine Genome arrays containing 24,123 probe sets, and the WHC of the muscle {\it longissimus dorsi} were determined in 74 animals. The data is publicly available at the Gene Expression Omnibus (accession GSE10204).

Among other analyses, \citet{ponsuksili2008} classified the probes as being associated or not to the WHC in the following way: A probe was declared associated to the WHC (AP) when the absolute value of the Pearson correlation coefficient between the WHC and the logarithm of the expression intensity of the probe exceeded 0.37 and the p-value for testing the significance of that Pearson correlation coefficient was (after accounting for multiple comparisons) smaller than 0.001 (corresponding to a q-value smaller than 0.004). Using this criterion 1,279 probes were classified as AP. We characterize here the distribution of the AP probes along the co-expression network.

\begin{figure}[!ht]
\centering
	\scalebox{.60}{\includegraphics{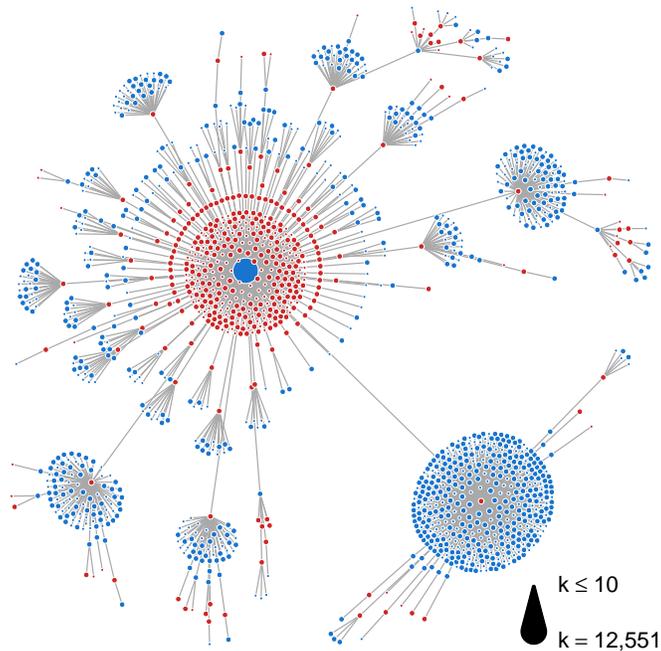}}
\caption{Co-expression network of Example 2 using the cluster representation.
Each point represents a cluster, which size is proportional to the number of probes in the cluster. Clusters with predominance of probes associated to the muscle water holding capacity are marked in red and the others in blue.}
\label{fig:GSE10204}
\end{figure}

The co-expression network of the 24,123 probes, inferred as the decomposable graphical model with minimum BIC, presented a single connected component with 34,842 edges. The highest vertex's degree was 69 and around 37\% of the vertices were leaves. Applying the proposed clustering procedure resulted in 1,521 clusters; 451 classified as having predominantly AP. The largest cluster (with 12,551 probes) had predominantly probes not associated to the WHC. Examining the distribution of the uncertainty index $\rho$ of the AP (see Figure~\ref{fig:rho}) it turns out that in this example there is a large amount of probes with low level of uncertainty. This is also apparent from the general structure of the network displayed in Figure~\ref{fig:GSE10204}.


The assumption of decomposability of the model was verified in a subset of the probes. We selected the 60 probes with highest variance (40 non-AP and 20 AP), calculated the likelihood function evaluated at the maximum for a complete model (using the inverse of the sample covariance matrix), and compared that with the likelihood function of an estimated decomposable graphical model. The likelihood ratio test yielded a p-value close to one. Note that this procedure was possible because we had 60 variables and 74 observation.

\FloatBarrier

\newpage
\noindent {\bf Example 3 - Intra-muscular fatness in pigs}\\

\noindent
This example arises from a study on muscle transcriptomic profiles in pigs described in \citet{canovas2010}. Samples from the muscle {\it gluteus medius} of 68 animals were collected and hybridized to Affymetrix GeneChip Porcine Genomic arrays with 4,299 probe sets. The animals were selected from two contrasting extreme phenotype groups: with high or with low intramuscular fat contents. \citet{canovas2010} found 1,060 probes presenting expression levels with significant differences between the two groups of animals. The data is publicly available at the Gene Expression Omnibus (accession GSE19275).

Here we inferred a co-expression network using the group of animals with high intramuscular fat contents and studied the distribution of the differentially expressed probes along this network. The decomposable graphical model estimated by minimizing the BIC presented one single connected component with 5,438 edges; the highest vertex's degree was 42 and around 38\% of the vertices were leaves.

\begin{figure}[!ht]
\centering
	\scalebox{.65}{\includegraphics{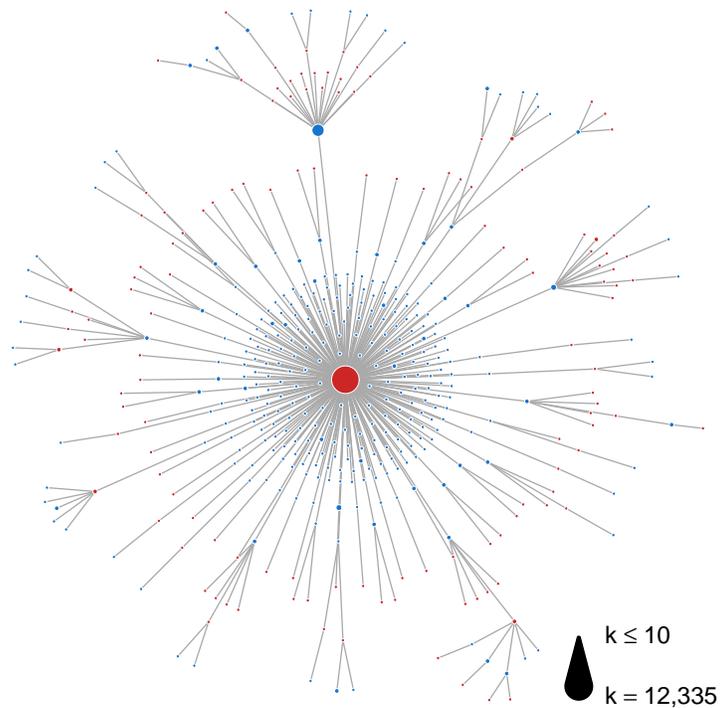}}
\caption{Co-expression network of Example 3. Each point represents a cluster, which size is proportional to the number of probes in the cluster. Clusters with predominance of differentially expressed probes are marked in red and the others in blue.}
\label{fig:GSE19275}
\end{figure}

\begin{figure}[!ht]
\centering
    \scalebox{.45}{\includegraphics{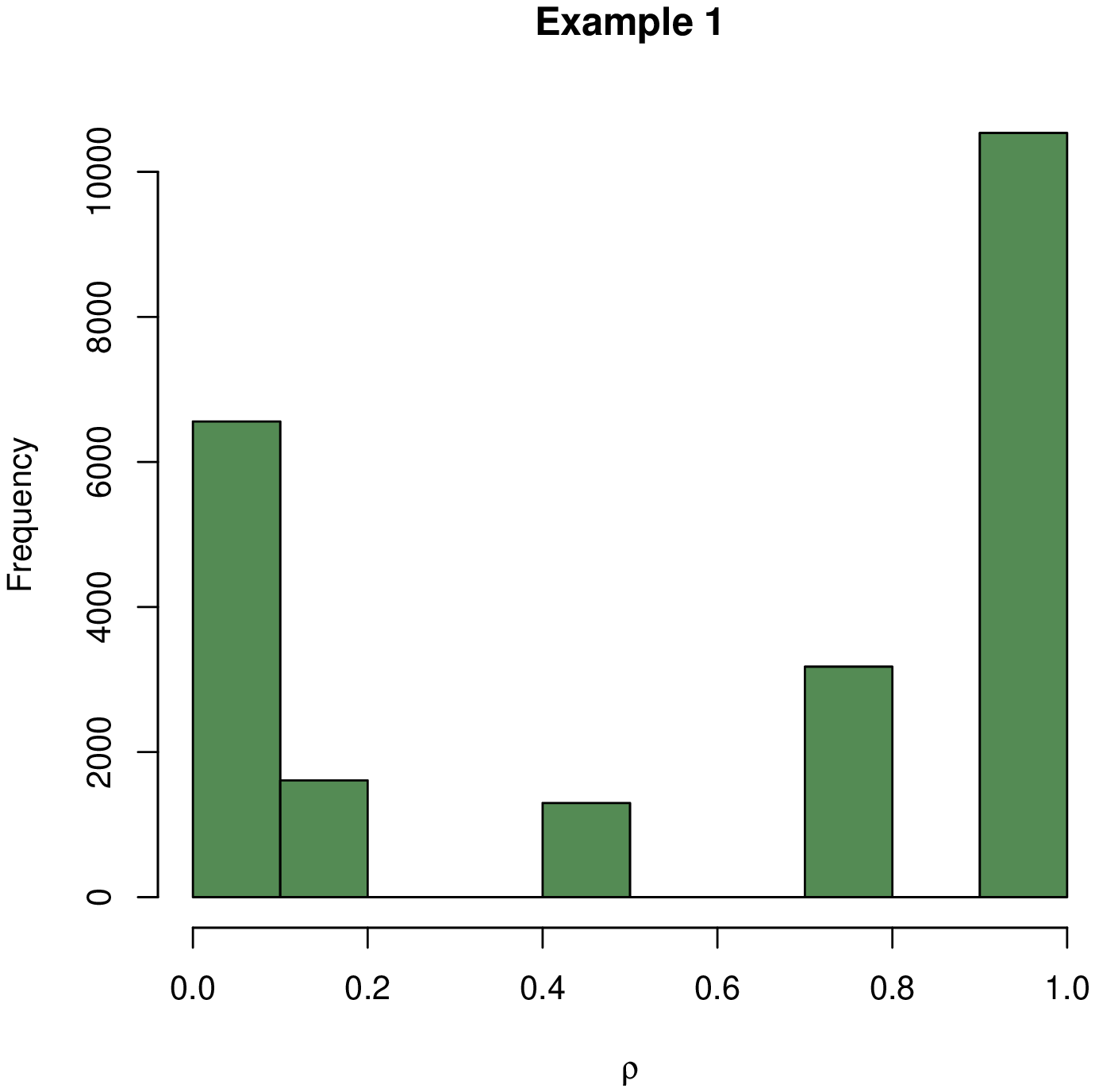}}
    \scalebox{.45}{\includegraphics{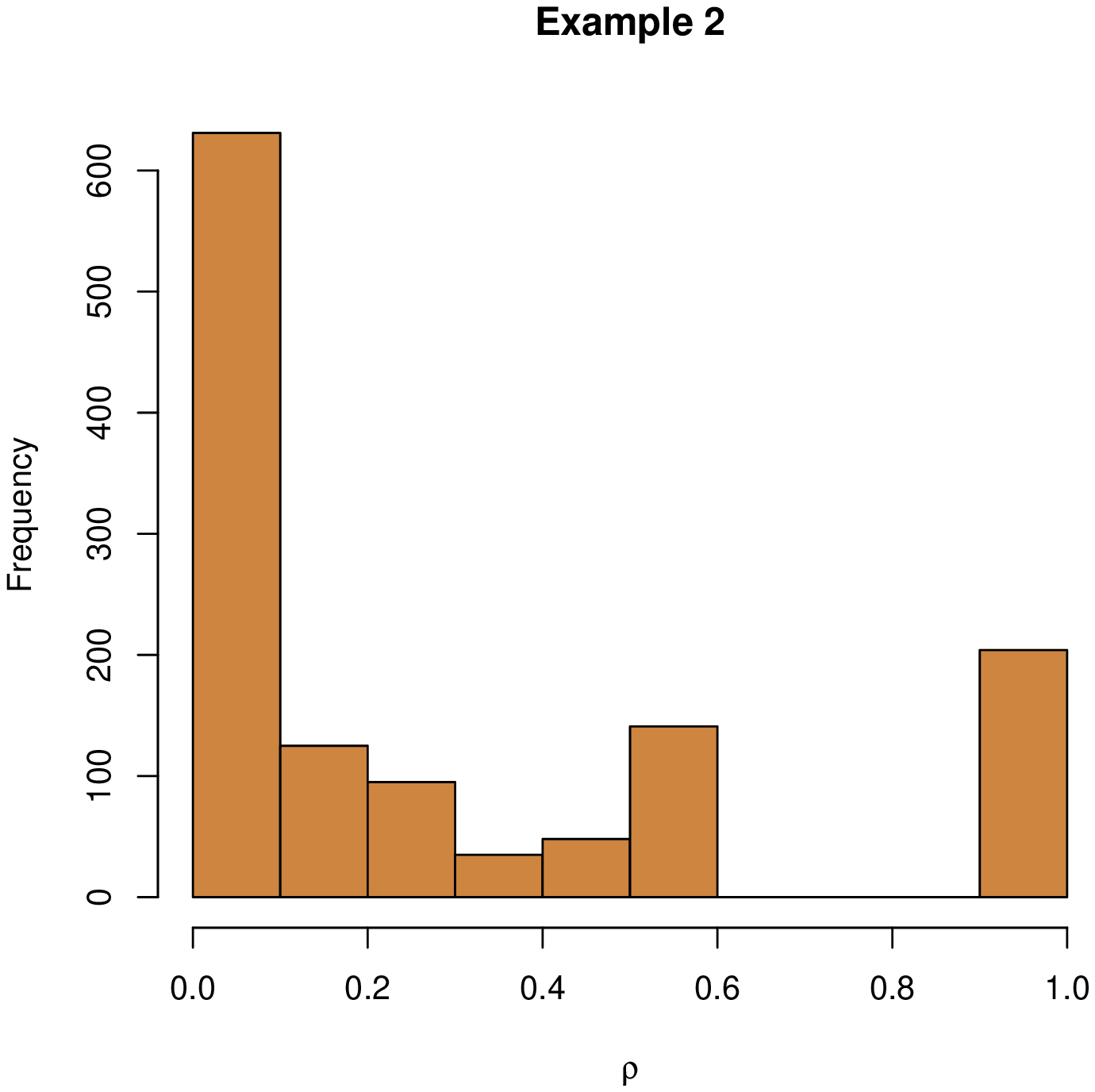}} \\
    \scalebox{.45}{\includegraphics{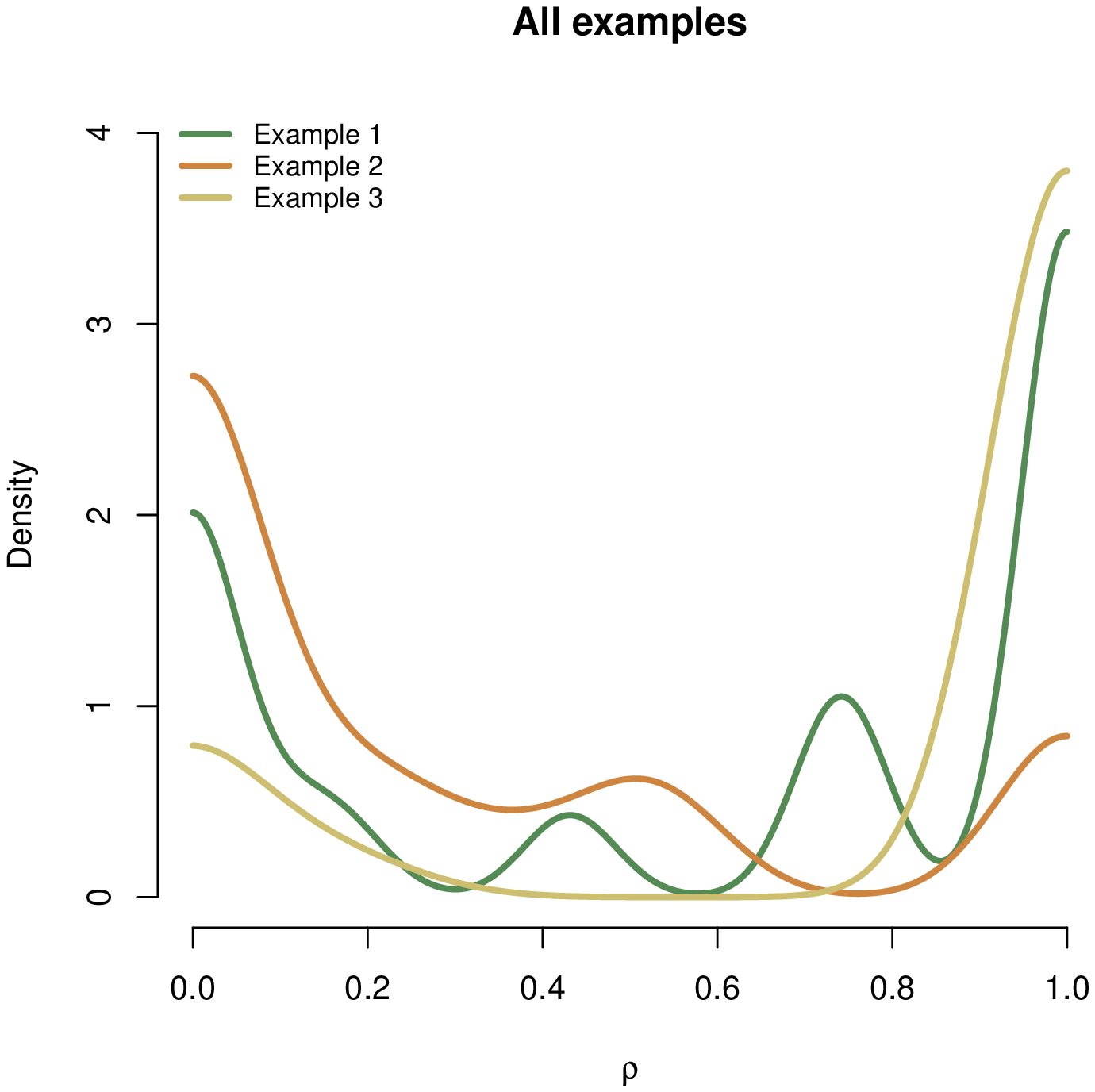}}
	\scalebox{.45}{\includegraphics{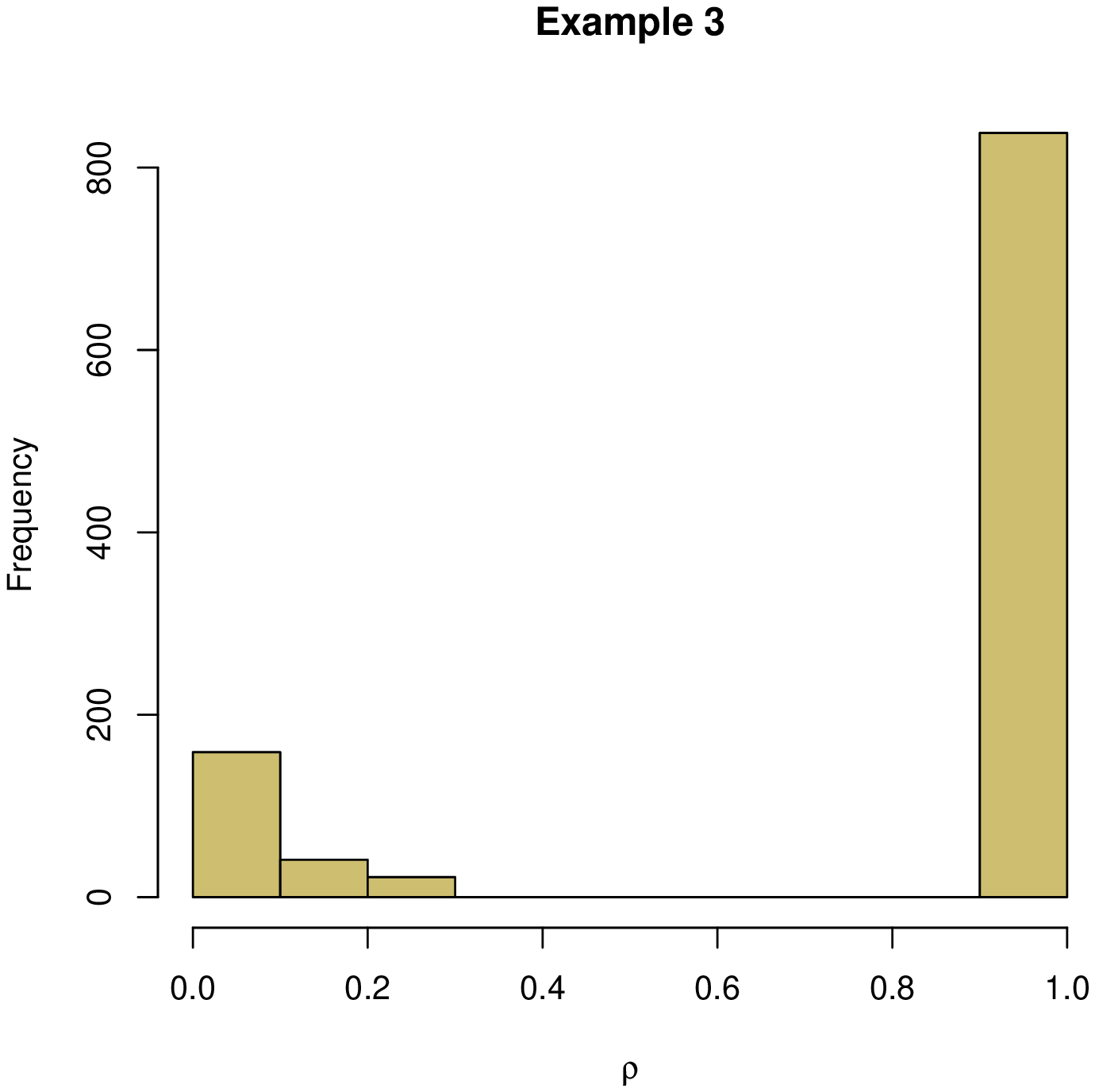}} 	
\caption{Histogram and density estimation (Gaussian kernel estimate, botton left plot) of the uncertainty coefficient ($\rho$) for the three examples.}
\label{fig:rho}
\end{figure}

Using our clustering procedure we obtained a representation of the network with 448 clusters, 132 of them presenting predominance of differentially expressed probes. This representation, displayed in Figure~\ref{fig:GSE19275}, presented a large central cluster (with 838 differentially expressed probes and in total 1,527 probes, \ie around 79\% of the differentially expressed probes and 36\% of the total of probes). The histogram of the measure $\rho$, displayed in Figure~\ref{fig:rho}, shows that a large proportion of the differentially expressed probes presented a high level of uncertainty, which contrasts strongly with the scenario in Example 2.

\vspace{1cm}

\FloatBarrier

\newpage

\FloatBarrier


\newpage

\FloatBarrier

\subsection{A short simulation study}
\label{subsec:simulation}
\begin{figure}[!ht]
\centering
	\scalebox{.45}{\includegraphics{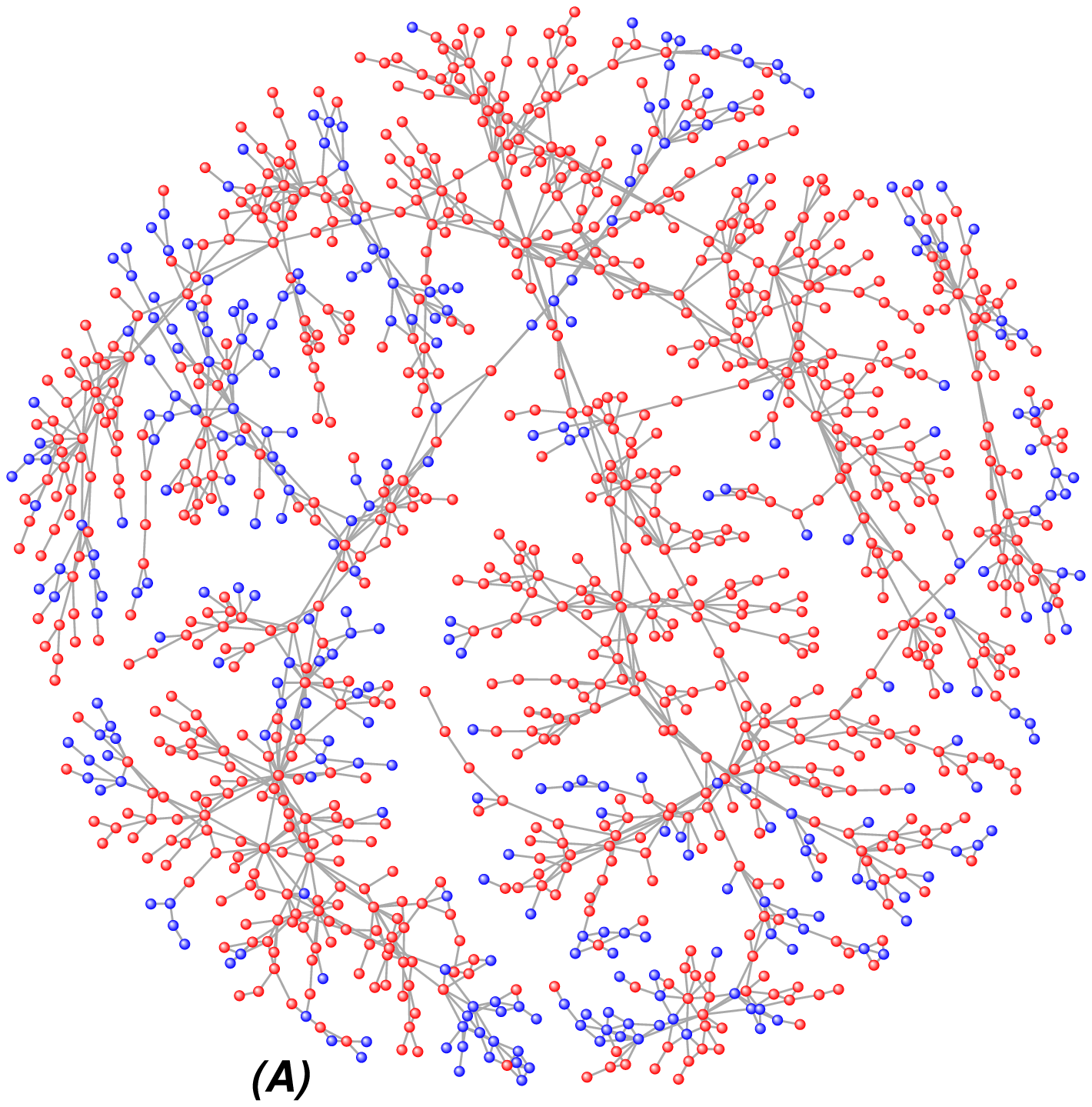}}
	\scalebox{.45}{\includegraphics{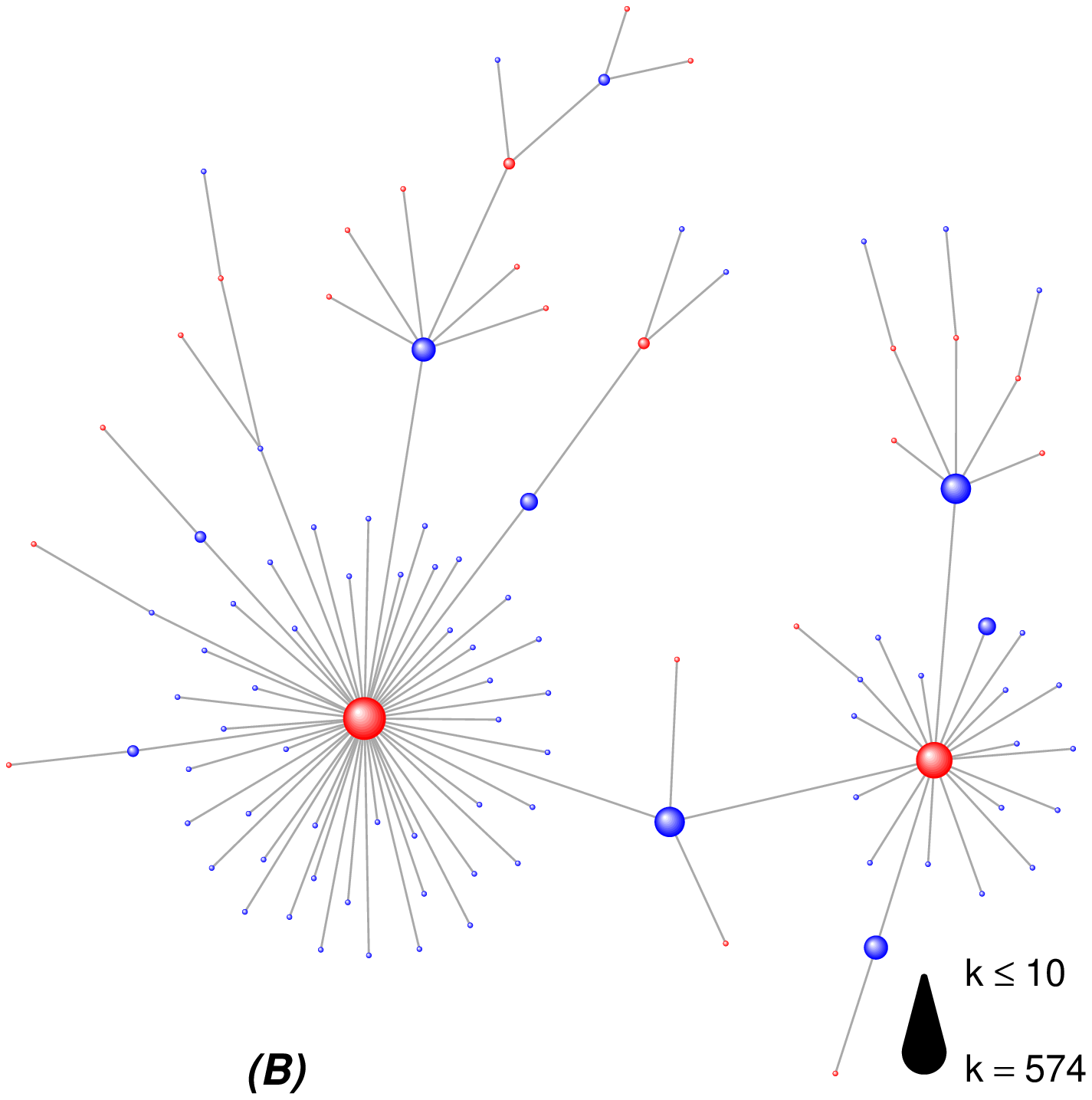}}
\caption{Co-expression network of the reference model constructed with the 1,200 probes with the largest variance in the data of Example 1. (A) Representation of the estimated decomposable graphical model; each dot represents a probe (differentially expressed in red, and non-differentially expressed in blue). (B) Representation based on the clustering procedure; each dot represents a cluster, which size is proportional to the number of probes in the cluster; clusters with predominance of differentially expressed probes are marked in red and the others in blue.}
\label{fig:simulOrig}
\end{figure}

The performance of the proposed cluster-based representation of networks was evaluated using the simulation study described below. First we constructed a reference data set by selecting the 1,200 probes with the highest variance in the data of Example 1 (Human healthy and diseased gingival tissues). This data set was then used to estimate a co-expression network following the procedure described in section \ref{subsec:basic} (see Figure~\ref{fig:simulOrig}). %
Next we estimated the covariance matrix of this decomposable graphical model by
$$
 \hat{\bSigma}=\left [ n\left\{\sum_{C\in{\cal C}}\left[(ssd_C)^{-1}\right]-\sum_{S\in{\cal S}}\nu(S)\left[(ssd_S)^{-1}\right]\right\}\right ]^{-1}\, .
$$
Here $n$ is the sample size, ${\cal C}$ is the set of cliques and ${\cal S}$ is the set of separators with multiplicity $\nu$ in a perfect sequence and, for a given set of vertices $A$, $ssd_A$ denotes $X_A^tX_A-X_A^t\vI_n\vI_n^tX_A/n$. We assume implicitly that $n>\max_{C\in{\cal C}}|C|$ \citep[see][]{lauritzen1996}. The covariance matrix $\hat{\bSigma}$ was then kept fixed and applied repeatedly to simulate multivariate normally distributed observations to be used in the performance evaluations.

\vspace{2cm}

\FloatBarrier

\newpage

\begin{figure}[!ht]
\centering
	\scalebox{.6}{\includegraphics{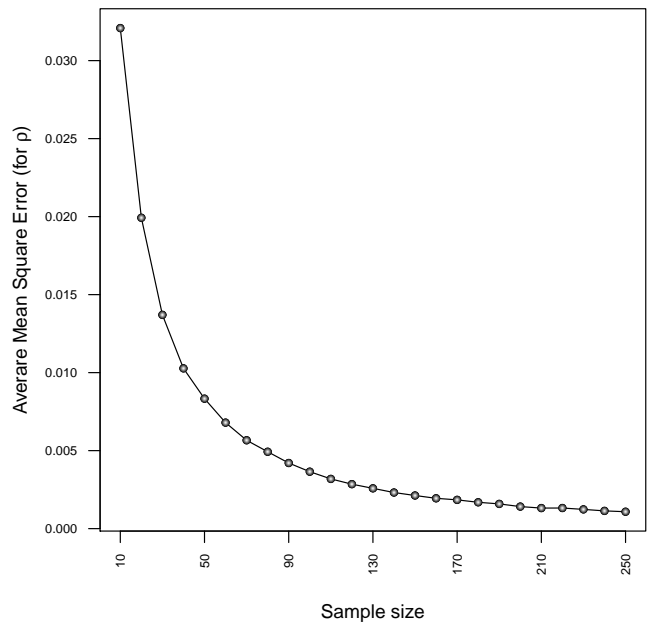}}
\caption{Monte Carlo estimate of the mean square error for the uncertainty index $\rho$ for different sample sizes.}
\label{fig:R2}
\end{figure}

The performance of the proposed procedure was then evaluated for 25 different sample sizes (from 10 to 250 increasing by 10). For each sample size we constructed 500 simulated data sets. For each of these data sets the uncertainty indices $\rho$ of the probes were calculated and compared to the uncertainty indices obtained from the reference data set.
Figure~\ref{fig:R2} displays the Monte Carlo estimates of the average mean square deviation of the uncertainty indices from the simulated reference. As expected the performance of the procedure improves substantially as the sample size increases.

\FloatBarrier


\section{Discussion}
\label{sec:discussion}
We presented arguments indicating that it is essential to take into account the intrinsic structure of interdependency of the expression levels of genes when discussing and interpreting the possible association of patterns of differential expression and the effect of a perturbation.
Our proposal is to use continuous graphical models to represent and infer this structure of interdependency. Since this representation uses conditional correlations to define the set of edges in the graph, instead of direct correlations as in relevance networks \citep{butte2003}, we avoid to introduce redundancy and to propagate spurious information in the representation of the network. Indeed, with this representation, two genes are directly connected to each other if, and only if, the expression level of each of them carries information on the expression level of the other in such a way that this information is not already contained in the expression levels of the rest of the genes in the network (see \cite{whittaker1990} chapter 4 for an argument based on information theory). Moreover, graphical models have suitable mathematical properties that are essential for our construction. One example is the separation principle \citep{whittaker1990,lauritzen1996}
which states that if two groups of variables, A and B, are separated by a third group of variables, C, then A and B are conditionally non-correlated given C. Here the expression ``group C separates A and B" means that every path connecting an element of A with an element of B necessarily contains an element of C. This basic property is essential for the interpretation of the relative position of a gene in the network. For instance, if a gene $v$ (or a group of genes) is placed in the central part of the network and separates the network in two parts, $A$ and $B$, then the knowledge of the values of the expression of $v$ renders the two branches of the network independent. This means that the expression level of $v$ carries all the information that the expression levels of all the genes in $A$ carry on the expression levels of the genes in $B$ (and vice-versa). Clearly, if the separation principle does not hold, then the position of a gene in the network looses interpretation.

Graphical models, in special Gaussian graphical models (or covariance selection models) have been known for a long period \citep[see][]{dempster1972}, however the use of these models to analyze large and complex gene-expression data has been limited. One of the reasons for this might be that throughput data on gene expression typically contains few observations (say 10 to 100) and a much larger number of variables (10 to 60 thousand genes), which makes the use of standard techniques for statistical inference for graphical models infeasible. For instance, the naive application of likelihood-based inference under the covariance selection model involves the inversion of the sample covariance matrix; however, it is known that if the number of variables exceeds the number of observations, then the sample covariance matrix is not invertible with probability one \citep{dykstra1970}.  A way to circumvent this problem is to base the inference on alternative methods and, at the same time, to restrict the class of graphical models considered. We based the inference of the co-expression network on the minimization of the BIC (Bayesian Information Criterion), since this method yields consistent and optimum estimates \citep[see][]{haughton1988}, although our methods could easily be adapted to other similar inference techniques  (\eg the minimization of the AIC (Akaike Information Criterion) or maximization of the entropy) and restricted the class models used to the Gaussian decomposable graphical models.

The class of decomposable graphical models has  suitable mathematical properties that we take advantage of in three different ways: First, these models can be separated in small components, the cliques, which act somehow independently \citep[see][]{golumbic1980} and which are used to construct the representation of the co-expression network needed for our methods. Secondly, these models allow for a special decomposition of the likelihood function in terms of the structure of cliques, which simplifies very much the calculations involved in some likelihood-based quantities such as the BIC \citep{lauritzen1996}. Thirdly, there exists a rather efficient algorithm already implemented for making statistical inference for high dimensional decomposable graphical models (the \emph{gRapHD} package described in \cite{abreu2010} and implemented in R, \citep{R}). The recognition of the edges that can be added to a triangulated graph in such a way that the new graph is also triangulated is a key operation in this algorithm. The computational resources required for this increase rapidly with the dimension of the graph if naive methods of direct verification are used, thus forming a bottle-neck of the algorithm. Therefore we improved the algorithm by implementing a more sophisticated technique to characterize decomposable graphical models which takes advantage of the structure of the cliques in decomposable models. The class of decomposable graphical models is the largest class of graphical models for which the mathematical properties mentioned above hold. Indeed, the existence of the referred decomposition of the models and of the likelihood quantities requires that the graph representing the model has a perfect enumeration, which is a necessary and sufficient condition for a graph to be triangulated (see theorem 4.3 in \cite{golumbic1980}).

An alternative form of inference of graphical models in general would be to minimize the BIC (or other penalized version of the likelihood function) by direct search or even by applying sophisticated Monte Carlo based methods for optimization (\eg simulated annealing as in used in \cite{Thomas2009}). These methods would not be feasible for large and complex graphical models for two reasons: First, the calculation of likelihood related quantities for non-decomposable graphical models with more vertices than observations is not an easy task; secondly, the number of possible graphical models for which the BIC should be evaluated increases rapidly with the number of vertices. Indeed, using the fact that the number of possible trees formed with $n$ vertices is $n^{n-2}$ (\cite{Godsiletal2000}, corollary 13.2.2), it is easy to see that the number of possible arbitrary graphs  that can be constructed even with relatively modest number of vertices are huge (\eg the number of possible trees with only 100 vertices is of the order of $10^{196}$).
Therefore, restrictions on the type of graphs must be introduced; for example \cite{edwards2010} considered only the sub-class of trees and forests while \cite{abreu2010} considered the class of decomposable graphs.

We claim that the restriction imposed by assuming decomposability of the graphical model does not seriously compromise the representation of the co-expression network. First, we can think on decomposable graphical models (and also spanning trees) as forming the skeleton of the (unrestricted) graphical model that adjusts the data. Chordless cycles of length larger than three might exist, but would be difficult to be detected with moderate sample sizes. Moreover, cycles of this type would not abound in a typical co-expression network since they would typically be associated with regulations of gene expressions in strict cascade in such way that the last gene in the cascade regulates the first. This cascade of gene regulation would be detected as a structure obtained by suppressing some edges from the cycle. For instance, making a local analysis of a subset of the probes in Example 2 we found using a likelihood ratio test that a saturated model (\ie the model associated with a complete graph) could be reduced to the estimated decomposable graphical model indicating absence of chordless cycles of length larger than three in this sub-graph. Techniques of local investigations in restricted sub-graphs could be implemented in the future as a model control or search for larger regulatory cycles.

The results of our simulation study indicated that the inference based on the minimization of the BIC performed well since the basic structure of the simulated network was recovered even for relatively small sample sizes. Moreover, the performance of the inference improved substantially when the sample size increased what is in accordance with the consistency of  the estimation via the minimization of the BIC predicted theoretically. Furthermore, the algorithm implemented is efficient enough for inferring large co-expression networks, as illustrated by Example 1 (54,675 vertices).

Once having established a reasonable model for the gene co-expression network, we turned to the question of characterizing the distribution of DEGs (differentially expressed genes) along this structure. As illustrated in Example 1 (see Figure~\ref{fig:perio54675a}~(A)), the naive approach of simply marking the DEGs does not reveal any pattern due to the complexity of the graph. However, we were able to propose a much more compact representation by exploring the internal structure characteristic of any decomposable graphical model - the structure of cliques - to determine regions of the co-expression network where the DEGs are over represented. This yields a second order graph, the cluster graph, with vertices representing clusters of genes located in cliques that are contiguous in the co-expression network. Each cluster is classified, by construction, in one of two categories:  presenting or not over representation of DEGs. The cluster graph can be thought as a graph with vertices with two colours. As illustrated in the example presented in Figure 2, the cluster graph yields a much more compact representation of the network as compared to the original co-expression network. The real example of diseased gingival tissues illustrates how drastic can this reduction in the complexity of the representation be in real gene expression data. Indeed, the complexity of the interdependency graph representing the original expression levels of the 54,675 probes in the Example 1 (diseased gingival tissues) is so high, that any pattern can be observed (see Figure~\ref{fig:perio54675a}~(A)). It is even not possible to realize in Figure~\ref{fig:perio54675a}~(A), that the differentially expressed probes (marked in a different colour) are not homogeneously distributed in the network. On the other hand, the cluster graph of Example 1 involves only  5,081 vertices (a reduction to less than 10\%) and clear patterns of  non-homogeneity in the distribution of DEGs can be observed  (see Figure~\ref{fig:perio54675a}~(B)). For instance, examining the cluster graph in details (see Figures~\ref{fig:perio54675b}~(B)-(D)) it is possible to identify very large clusters containing predominantly DEGs and located in central areas of the network. The cluster graph contains also many small clusters located in the periphery of the network and presenting excess of DEGs. These two types of clusters, we argue, are of very different nature. While the expression levels of genes located in large central clusters cannot be disentangled from each other, the genes in the small peripheral clusters act essentially as isolated from the rest of the network. These fundamental qualitative differences in the DEGs cannot be identified if we use the original co-expression network.

Once the co-expression network is stratified in non-overlapping contiguous regions (the clusters), it is natural to classify the DEGs genes according to the number of DEGs present in the cluster they belong to. The associations between the perturbation made in a differential expression study (e.g. treatments, types of tissue, etc) and the expression levels of genes belonging to a cluster with many other DEGs cannot necessarily be attributed to a single gene or even a reduced number of genes. Those genes are considered less informative, since the knowledge that they are differentially expressed does not reduce very much the uncertainty that we had before performing the experiment. On the other hand, knowing that a gene is differentially expressed adds more information if the gene is located in a cluster containing only few DEGs.  This principle is used when defining the information index $\rho$.  We expressed $\rho$ as  the negative  proportion of total number of DEGs represented by the number of DEGs in the cluster multiplied by the logarithm of this proportion, since this quantity resembles the classic entropy  used to represent uncertainty in information theory.

In conclusion, we devised a method to construct compact representations of co-expression networks that allows the identification of regions with high concentration of differentially expressed genes and genes with high informational content. We also presented an alternative method of inference of gene co-expression networks based on high dimensional decomposable graphical models and an efficient algorithm that allowed us to treat typical throughput data on gene expression using reasonable amounts of computational resources.

\section*{Acknowledgements}
G.C.G. Abreu was financed by SABRETRAIN Project, funded by the Marie Curie Host Fellowships for Early Stage Research Training, as part of the $6^{th}$ Framework Programme of the European Commission. R. Labouriau was partially supported by the project ``Metabolic programming in Foetal Life", Danish Research Agency, Ministry of Science Technology and Innovation.

\bibliographystyle{plainnat}


\end{document}